\documentclass[fleqn,usenatbib]{mnras}

\usepackage{newtxtext,newtxmath}
\usepackage{graphicx}
\usepackage[T1]{fontenc}
\usepackage{ae,aecompl}
\usepackage{lineno}

\usepackage{amsmath}	% Advanced maths commands

\usepackage{layout}

\usepackage{scalerel,stackengine}
\stackMath
\newcommand\reallywidehat[1]{%
\savestack{\tmpbox}{\stretchto{%
  \scaleto{%
    \scalerel*[\widthof{\ensuremath{#1}}]{\kern-.6pt\bigwedge\kern-.6pt}%
    {\rule[-\textheight/2]{1ex}{\textheight}}%WIDTH-LIMITED BIG WEDGE
  }{\textheight}% 
}{0.5ex}}%
\stackon[1pt]{#1}{\tmpbox}%
}
\newcommand{\Cn}{C_n^{(\alpha)}}

\newcommand{\vx}{\vec{x}}

\newcommand{\vr}{\vec{r}} %vector r
\usepackage[dvipsnames]{xcolor}

\newcommand{\kg}[1]{\textcolor{magenta}{\textbf{#1}}}

\title[Improved Anisotropic 3PCF]{Improving the Line of Sight for the Anisotropic 3-Point Correlation Function of Galaxies: Centroid and Unit-Vector-Average Methods Scaling as $\mathcal{O}(N^2)$}

\author[Garcia \& Slepian]{
Karolina Garcia$^{1}$\thanks{E-mail: karolina.garcia@ufl.edu (KG)} \&
Zachary Slepian$^{1,2}$\thanks{E-mail: zslepian@ufl.edu (ZS)} \\
$^{1}$Department of Astronomy, University of Florida, 211 Bryant Space Science Center, Gainesville, FL 32611, USA\\
$^{2}$Lawrence Berkeley National Laboratory, 1 Cyclotron Road, Berkeley, CA 94720, USA}

\date{Accepted XXX. Received YYY; in original form ZZZ} 

\pubyear{2020}

\begin{document}
\label{firstpage}
\pagerange{\pageref{firstpage}--\pageref{lastpage}}

\maketitle

\begin{abstract}
The 3-Point Correlation Function (3PCF), which measures correlations between triplets of galaxies, is a powerful tool for the current era of high-data volume, high-precision cosmology. It goes beyond the Gaussian cosmological perturbations probed by the 2-point correlation function, and includes late-time non-Gaussianities introduced by both nonlinear density field evolution and galaxy formation. The 3PCF also encodes information about peculiar velocities, which distort the observed positions of galaxies along the line of sight away from their true positions. To access this information, we must track the 3PCF’s dependence not only on each triangle’s shape, but also on its orientation with respect to the line of sight. Consequently, different choices for the line of sight will affect the measured 3PCF. Up to now, the line of sight has been taken as the direction to a single triplet member (STM), but which triplet member is used impacts the 3PCF by $\sim$20\% of the statistical error for a BOSS-like survey. For DESI (2019-24), which is 5$\times$ more precise, this would translate to $\sim$100\% of the statistical error, increasing the $total$ error bar by $\sim$40\%. We here propose a new method that is fully symmetric between the triplet members, and uses either the average of the three galaxy position vectors (which we show points to the triangle centroid), or the average of their unit (direction) vectors. We prove that these two methods are equivalent to $\mathcal{O}(\theta^2)$, where $\theta$ is the angle subtended at the observer by any triangle side. Naively, these approaches would seem to require triplet counting, scaling as $N^3$, with $N$ the number of objects in the survey. By harnessing the solid harmonic shift theorem, we here show how these methods can be evaluated scaling as $N^2$. We expect that they can be used to make a robust, systematics-free measurement of the anisotropic 3PCF of upcoming redshift surveys such as DESI. So doing will in turn open an additional channel to constrain the growth rate of structure and thereby learn the matter density as well as test the theory of gravity.
\end{abstract}


%%%%%%%%%%%%%%%%%%%%%%%%%%%%%%%%%%%%%%%%%%%%%%%%%%

%%%%%%%%%%%%%%%%% BODY OF PAPER %%%%%%%%%%%%%%%%%%

\section{Introduction}
\label{sec:intro}
%Inflation + generation of GRF perturbationos
During the first $10^{-35}$ seconds, the universe experienced a period of exponential expansion known as inflation. The field driving this then decayed into radiation and matter, producing density fluctuations standardly taken to be a Gaussian Random Field (GRF).\footnote{GRF means the real and imaginary parts of the field expressed in Fourier space are drawn from a Gaussian whose variance is the power spectrum, and the complex phase is uniform. The initial density field is taken to be a GRF modulo very small possible additional contributions known as primordial non-Gaussianity (PNG).} For a GRF, the 2-point correlation function (2PCF), measuring the excess probability over random of finding a pair of points with given density fluctuations separated by a given distance, captures all the information.
%Nonlinear structure formation and higher-oorder correlationos
However, over the Universe's subsequent evolution, gravitational interactions led to nonlinear structure formation, inducing additional correlation in the density fluctuation field. Hence this field began to deviate from a GRF, and higher-order correlation functions arose. In particular, nonlinear evolution of matter under gravity produces a 3-Point Correlation Function (3PCF) for the matter, where the 3PCF characterizes the excess probability over random of observing three density fluctuation values on a given triangle configuration (e.g. \citealt{Bernardeau:2002}). Furthermore, galaxies do not perfectly trace the matter (galaxy biasing, e.g. \citealt{Desjacques:2018}), which induces additional higher-order correlations, including contributions to the 3PCF (e.g. \citealt{Gaztanaga_1994}, \citealt{Jing_2004}, \citealt{Guo_2015}, and \citealt{SE_3PCF_model}; also see e.g. \citealt{Scoccimarro_1999} or \citealt{Gil_Marin_2015} for discussion of biasing in the 3PCF's Fourier space analog the bispectrum).

%Uses of 3PCF
Since the 3PCF stems from both nonlinear matter evolution and galaxy biasing, on  large scales it can be used in conjunction with the 2PCF to disentangle matter clustering from galaxy bias.\footnote{Matter clustering can be summarized as $\sigma_8$, the rms amplitude of the density fluctuation field on $8$ Mpc$/h$ spheres, and the traditional argument is that $2PCF^2/3PCF$ isolates one power of the linear bias $b_1$, where in the simplest models the galaxy density fluctuation $\delta_{\rm g} = b_1 \delta_{\rm m}$ with $b_1$ the linear bias and $\delta_{\rm m}$ the matter density fluctuation; e.g. \cite{Fry:1993}.} It can also be used to investigate in more detail how galaxies trace the matter, since it is sensitive to more complicated models of galaxy biasing (e.g. scaling as the matter field squared, or as the matter field's tidal tensor \citep{McDonald:2009, Chan_2012}, or as the baryon-dark matter relative velocity \citep{Yoo_2011, SE_RV_sig, SE_RV_constraint}) at leading order, in contrast to the 2PCF which contains these terms only at sub-leading order. On even larger scales, the 3PCF contains Baryon Acoustic Oscillations (BAO) and so measuring it can offer a standard ruler by which to gauge the cosmic expansion history and in turn constrain dark energy \citep{SE_3PCF_BAO}; for the analogous method using the bispectrum see \cite{Pearson_2018}.

%Introduce RSD
However, as is true for the 2PCF as well, the 3PCF is affected by the fact that galaxies' true, 3D positions are unknown. Their peculiar velocities introduce a component to the observed redshift beyond what would come from co-movement with the background expansion alone. This in turn affects the distance inferred for the galaxy. The distortions induced by these velocities on the observed map of 3D galaxy positions are termed Redshift-Space Distortions (RSD).

Physically, the peculiar velocities producing RSD stem from two factors:
\begin{enumerate}
    \item the large-scale density field, in which potential wells attract galaxies, and make them appear bluer when their additional recessional velocity is pointing to the observer, and redder when it points away from them. The galaxy distribution appears squashed along the line of sight due to this effect, referred as the Kaiser effect \citep{Kaiser1987}; and
    \item the motion of satellite galaxies within clusters, which cause random motions on galaxies in smaller scales. Since in this case the velocities are random inside the clusters, and we only measure the radial contribution, structures on cluster scales will seem to have an elongated shape, referred to as Fingers of God (FOG; \citealt{Jackson1972}).
\end{enumerate}
No model for the redshift-space 3PCF including both of these effects yet exists, though a number of works investigate RSD in the bispectrum, the 3PCF's Fourier-space analog. \cite{Scoccimarro_1999} uses Eulerian Standard Perturbation Theory (SPT) and incorporates effect (i), and \cite{Rampf_2012} use Lagrangian Perturbation Theory (LPT) and include re-summed contributions which correspond to higher-order SPT terms in the redshift-space bispectrum, but the RSD treatment still includes just (i). Developments of the \cite{Scoccimarro_1999} model by \cite{Gil_Marin_2015} also include effect (ii), and this model was further extended to the case of Primordial Non-Gaussianity (PNG) by \cite{Tellarini_2016}.

\iffalse
{\bf KG: please check over what Rampf and Wong paper does; please check how they include RSD and how it connects to scoccimarro and your effects (i) and (ii)}
\kg{They finish with: "Note that, unlike the case of the real-space bispectrum, redshift-space distortions induce for Bs a dependence on the direction of the constituent wavevectors. We leave the numerical evaluation of equation (4.16) for a future project." So I dont think RSD are included.}

{\bf KG: we need to check what Senatore paper does}

{\bf KG: look at Smith paper \cite{Smith_2008}, https://arxiv.org/pdf/0712.0017.pdf--this is on smaller scales}

{\bf KG: we need to add in Gualdi paper forecasting anisotropic bispectrum constraints, but probably this will come in a bit later; https://arxiv.org/abs/2003.12075, \cite{Gualdi_2020}}

{\bf KG: simialr to above, regarding forecasts, we want to include this paper: https://arxiv.org/abs/1807.07076, \cite{Yankelevich_2018}}

{\bf KG: we may want to look at this paper too :Large-scale bias in the Universe – II. Redshift-space bispectrum by Licia Verde}

{\bf KG: we need to find a possible senatore paper on this}
\fi

While some of the above models characterize the dependence of the bispectrum on the triangle's orientation with respect to the line of sight (e.g. \citealt{Scoccimarro_1999}, \citealt{Rampf_2012}), there have been only a very few works showing how to measure the anisotropic bispectrum or 3PCF; we discuss those that exist in \S\ref{sec:Sugi_Slep}. Yet the time is ripe for further developing such algorithms. Using an algorithm presented in the same work, \cite{Sugiyama_2018} recently reported the first evidence for an anisotropic signal in the 3PCF’s Fourier-space analog, the bispectrum, and using mock catalogs, \cite{sugiyama2020selfconsistent} investigated self-consistent fitting of the anisotropic 2PCF and anisotropic 3PCF. Moreover, a spate of recent works have forecast that measuring anisotropic bispectrum of 3PCF offers significant gains on cosmological parameters. \cite{Gagrani_2017}  forecasts that anisotropic bispectrum offers a factor of 3 improvement on the logarithmic growth rate $f$ relative to 2PCF alone. Even if we treat linear bias and the clustering normalization ($\sigma_8$) as fully unknown (a highly conservative choice), \cite{Gualdi_2020} forecasts that 3PCF offers a 30\% improvement over 2PCF alone on each of these parameters (including $f$). Thus, measuring the anisotropic 3PCF, and doing so with both high accuracy and high precision, is an important way to extend the reach of upcoming surveys such as Dark Energy Spectroscopic Instrument (DESI; \citealt{levi_desi}).

However, the line of sight used to a given triplet of galaxies is an important piece of any such effort. All previous works have used the line of sight as given by one of the three triplet members, as discussed in more detail in \S\ref{sec:Sugi_Slep}. We term this approach ``single-triplet member,'' or STM for short. However, \cite{Sugiyama_2018} shows that the anisotropic bispectrum changes by as much as 20\% relative to its statistical errorbars as one cycles from using one galaxy as the line of sight to another to the third. This work was done for a Sloan Digital Sky Survey (SDSS) Baryon Oscillation Spectroscopic Survey (BOSS)-like sample, and the contribution of such an error would be only a 2\% increase in the total error budget. However, DESI will have roughly 5$\times$ the precision of BOSS, and hence a 20\% effect relative to BOSS's statistical errorbars would be a 100\% one relative to DESI's. Thus, the change as one shifts from one galaxy to another in a triplet to define the line of sight would inflate DESI's anisotropic bispectrum errorbars by as much as a factor of $\sqrt{2}$, meaning a 40\% increase in the total error budget. While one might think that averaging over the three choices for line of sight on each galaxy triplet would cause some of this error to cancel, much as happens for the anisotropic 2PCF when the line of sight is taken to be a single galaxy pair member (see \citealt{SE_wa_FT} for reasons) we further outline in \S\ref{sec:Sugi_Slep}, this cancellation does not occur.

Hence, it is worth considering if an estimator that uses a more symmetric definition of the line of sight to a triangle (rather than just choosing one triangle member at a time) can be developed. Furthermore, given the typical computational expense of 3PCF, it is worth seeking an estimator that harnesses the algorithmic innovations of \cite{SE_3pt_alg}, \cite{SE_aniso_3PCF} or \cite{Sugiyama_2018}, all of which exploit spherical harmonics to factorize the 3PCF or bispectrum calculations. \cite{Scoccimarro_2015} also has a fast algorithm but, as we will detail further in \S\ref{sec:Sugi_Slep}, it does not capture the full anisotropic information. 

In this work, we develop a fully symmetric approach to defining the line of sight to a galaxy triplet. We investigate two choices: first, a straight average of the three absolute position vectors of the galaxies in the observer's frame, and second, an average of their direction vectors (i.e. make each position vector into a unit vector) in the observer's frame. We prove that the first choice actually passes through the centroid of the triangle, and hence term that method the ``centroid method.'' We term the second method the "unit-vector average" method. We also show that these two choices differ from each other only at $\mathcal{O}(\theta^2)$, where $\theta$ is the ratio of the typical triangle side to the distance of the triangle from the observer. $\theta$ is essentially the angle each side subtends at the observer, and for small $\theta$, we are in the flat sky limit. We note that both the above choices of lines of sight are fully symmetric under interchange of the triplet members with each other, unlike the ``single-triplet member'' estimators previously used.

Most critically, in this work we show how the anisotropic 3PCF using the above definitions of the line sight can be evaluated scaling as $N^2$, where $N$ is the number of objects in the survey. To do so, we exploit the Solid Harmonic Shift Theorem to develop a series expansion for spherical harmonics of our line of sight in terms of spherical harmonics of each triplet member's position vector. Our expressions give the exact result for evaluating these improved lines of sight to arbitrary precision. However, the computational cost does rise slightly. Nonetheless, to obtain the leading-order correction that goes beyond STM requires only a very modest increase in computational work. We present explicit expressions for the leading-order correction terms needed.

\iffalse
{\bf ZS: then we will briefly discuss anisotropic bispec algo of Sc as below, then move to Sugiyama and Slepian and acutlaly say we will gvie more detail on these in a separae section anywa.}
i) Scocimarro https://arxiv.org/abs/1506.02729,\cite{Scoccimarro_2015} Slepian Eisenstein 2018\\
ii) Sugiyama 2018\\
iii) recent sugiayma paper: https://arxiv.org/pdf/2010.06179.pdf
measured anisotropic bispectrum on BOSS mocks\\

Discussion of ST line of sight can elead into this binning/non-cancellation discussion.

Discuss binning and non-cancellation. {\bf Has to go in intro because it actually is the reason our symmetrization is important.}

Paragraph saying we wanna do better than that and then explain briefly our method.
{\bf KG: can you take a stab at this ``explaining our method paragraph": basically the improtant point is we don't wnat to triplet count so we have to factorize the weight $Y_{\ell m}^*$ so that you can split up integrals}
\fi

This paper is laid out as follows. In \S\ref{sec:Sugi_Slep} we summarize previous algorithms for the anisotropic 3PCF and bispectrum. In \S\ref{sec:basis} we present the basis used in this work. In \S\ref{sec:los} we discuss our choices of line of sight and prove both that the average of position vectors gives the triangle centroid, and that this method and the unit-vector-average method agree at $\mathcal{O}(\theta^2)$. In \S\ref{sec:factorization} we show how to evaluate our basis in a factorized fashion that permits an $\mathcal{O}(N^2)$ scaling. \S\ref{sec:leading} gives explicit expressions for the leading-order terms required by our method to go beyond STM. \S\ref{sec:conclusions} concludes. In Appendix \ref{app:app1} we derive the core expansions, of spherical harmonics of a sum of vectors into spherical harmonics of single vectors, that this work harnesses. This uses the solid harmonic addition theorem, and to offer the reader some intuition on this useful mathematical result, in Appendix \ref{app:app2} we provide an explicit proof of the low-$\ell$ cases of the theorem using Cartesian forms of the spherical harmonics. Finally, in Appendix \ref{app:app3} we show how to turn a product of spherical harmonics of the same argument into a sum over single harmonics, a result we require in the work.

\section{Previous Work on the Anisotropic Bispectrum and 3PCF}
\label{sec:Sugi_Slep}
We here outline three previous approaches to the anisotropic bispectrum and 3PCF. In this work, we build on the third of these approaches.  The second is actually exactly equivalent to the third in terms of the means of evaluating; the bases of the second and third are simply related by a linear transformation. Our discussion of the first is for the sake of completeness, and to fully explore the issue of how rotating among the three triplet members to define the line of sight enters previous methods.  

\subsection{Scoccimarro}
\label{subsec:Scocc}
We here discuss three previous works that presented algorithms for computing the anisotropic 3PCF or bispectrum. First, \cite{Scoccimarro_2015} developed a method to obtain multipole moments of the bispectrum. We first briefly articulate the parametrization used in that work, as understanding it is necessary to then follow its treatment of the line of sight, denoted $\hat{n}$. \cite{Scoccimarro_2015} parameterizes the bispectrum by the angle cosine $\mu \equiv \hat{k}_1 \cdot \hat{n}$ of the largest wave-vector, defined to be $\vec{k}_1$, to the line of sight. A further parameter is the azimuthal angle $\omega$ of the second-largest, defined to be $\vec{k}_2$, about $\vec{k}_1$. This parametrization is necessary because, at fixed internal triangle angle, the orientation of $\vec{k}_2$ is not fully independent from that of $\vec{k}_1$.  At fixed $\mu \equiv \hat{k}_1 \cdot \hat{n}$, $\vec{k}_1$ lives on a cone about the line of sight, and at fixed $\vec{k}_1$, $\vec{k}_2$ lives on a cone about $\vec{k}_1$ with opening angle cosine $\hat{k}_1 \cdot \hat{k}_2$.

To parametrize the triangle itself, \cite{Scoccimarro_2015} then uses $k_1, k_2, k_3$, the three triangle side-lengths in Fourier space. These fully characterize the triangle; we term them ``internal'' parameters, and term $\mu$ and $\omega$ ``external.''  \cite{Scoccimarro_2015} then averages over rotations of $\vec{k}_2$ about $\vec{k}_1$, reducing the anisotropic bispectrum to a 4D function of $k_1, k_2, k_3$ and $\mu$ whose $\mu$-dependence can then be expanded in Legendre polynomials. That work then bins the bispectrum by $k_1, k_2, k_3$. On each such bin, it records multipoles $\ell = 0$ and $2$ with respect to $\mu$. 

In detail, this is done by taking a ``local'' bispectrum estimate about a point $\vec{x}$, initially taken to be the centroid of a triangle formed by three galaxies in configuration space. Hence, the bispectrum is formed as an average over contributions taken on a galaxy-triplet-by-galaxy-triplet-basis. One then averages over all triplets by integrating over $\vec{x}$. \cite{Scoccimarro_2015} suggests using $\vec{x}$ as defining the line of sight to the galaxy triplet, and forming multipoles with respect to $\hat{k}_1 \cdot \hat{x}$. To accelerate the estimator, \cite{Scoccimarro_2015} then replaces $\vec{x}$ with $\vec{x}_i$ , where $\vec{x}_i$ is the position vector of a single triplet member. Consequently, we term this a ```single triplet member'' (STM) method. 

Overall, each galaxy triplet will contribute to a given $k_1, k_2, k_3$ bin three times: once with $\hat{x}_1$, once with $\hat{x}_2$, and once with $\hat{x}_3$ as the line of sight for the multipoles. This point is important because it connects to the analogous procedure for the anisotropic 2PCF, known as the ``single pair member'' (SPM) line of sight or the ``Yamamoto approximation'' \citep{Yamamoto_2006}. This method for the 2PCF results in cancellation of the difference between it and the angle bisector or separation midpoint (for the 2PCF) up to order $\theta^2$ \citep{SE_aniso_wa}. $\theta$ is the opening angle of the triangle formed by the observer and the galaxy pair. In particular, for the 2PCF, this cancellation occurs because each galaxy pair contributes twice to whatever separation bin it enters, once with each pair member's position defining the line of sight. The fact that the \cite{Scoccimarro_2015} approach allows each triplet member to define the line of sight in turn and co-add into one $k_1, k_2, k_3$ bin might suggest that an analogous cancellation to that in the 2PCF applies. Specifically, we might ask if the \cite{Scoccimarro_2015} STM method agrees with using the line of sight to e.g. the triangle centroid up to $\mathcal{O}(\theta^2)$. However, we have not been able to prove such a result.

%2) how he loses information by averaging over k2 about k1

%3) how he bins and how his estimates enter the bin

%Slepian and Eisenstein
\subsection{Slepian \& Eisenstein}
\cite{SE_aniso_3PCF} presented a method to compute the spherical harmonic decomposition of the anisotropic 3PCF scaling as $N^2$. In their earlier work \cite{SE_3pt_alg}, they showed how to evaluate the isotropic (averaged over rotations of the triangles) 3PCF in the basis of Legendre polynomials for the dependence on triangle opening angle by taking local spherical harmonic decompositions. One sat at a given galaxy (the primary), expanded the density into spherical harmonics on spherical shells around this primary, and then considered combinations of the harmonic coefficients on pairs of bins. The harmonic expansion about each galaxy scales as $N$ (more technically, $nV_{R_{\rm max}}$ where $n$ is the survey number density and $V_{R_{\rm max}}$ is the volume of a sphere of radius $R_{\rm max}$ with $R_{\rm max}$ the maximal scale out to which correlations are measured). Combining the coefficients on pairs of bins scales as $N_{\rm bins}^2$, which is modest given that typically 10-20 bins are used; critically, it is also independent of the number of objects in the survey. The local isotropy (averaging the triangles over rotations about each primary) means that only equal-total-angular momentum combinations of spherical harmonics can enter; in detail, one wants zero-total-angular momentum combinations of coefficients about the primary because they are the only ones that can survive under rotation-averaging. More detailed explanation of this point, for combining an arbitrary number of coefficients, is in \cite{Cahn_2020}. 

When one wishes to track anisotropy, one can form more general combinations of spherical harmonic coefficients about the primary. In particular, one can form combinations with any $\ell + \ell'$ where the sum is even (because parity-symmetry is still preserved). If one chooses that the $z$-axis of the local spherical harmonic expansion is the line of sight, then only harmonic coefficient combinations with $m = -m'$ appear, because there is still symmetry under rotations around the line of sight. Making the line of sight the $z$-axis means that rotations around it are purely in $\phi$, hence causing the $\phi$-dependent part of the two spherical harmonics involved, $\exp[i (m -m')\phi]$, to have a selection rule that $m-m'= 0$.

If the primary galaxy is used to define the line of sight, at each primary, one can rotate into the correct frame, perform the spherical harmonic decomposition, and then compute the combinations of coefficients on each bin pair. Hence, the algorithm still scales as $N^2$ just as the \cite{SE_3pt_alg} isotropic 3PCF algorithm does. There is some additional computational expense of performing the rotation at every primary, but this can be done very efficiently with matrix multiplication. \cite{Friesen_2017} presents a highly efficient implementation of this algorithm. Furthermore, \cite{SE_aniso_3PCF} shows that this algorithm can be evaluated using Fourier Transforms (FTs), much as the isotropic one could be \citep{SE_3pt_FT}. To do so, one obtains the spherical harmonic coefficients in one ``global'' basis around all primaries using convolutions evaluated by Fast FTs, and then rotates after the fact by using Wigner D-matrices defined at each primary to appropriately ``locally'' rotate the harmonic coefficients that had been computed about each primary in the ``global'' basis. This latter version of the algorithm has not yet been implemented but would scale as $N_{\rm g} \log N_{\rm g}$, with $N_{\rm g}$ the number of grid points used for the FT.

An important point for the current work relates to the way that galaxy contributions are accumulated to each bin. In the \cite{SE_aniso_3PCF} algorithm, as in its isotropic sibling, one vertex of the triangle is chosen as the origin (where the ``primary'' sits). The 3PCF or anisotropic 3PCF is then reported as radial coefficients on bins in the side lengths $r_1$ and $r_2$ extending from that primary. However, as the algorithm cycles over a survey, each triangle of galaxies is counted three times, once with each vertex. Hence, each triplet member does get its chance to define the line of sight to a given triangle of galaxies. However, the contributions from each line of sight are accumulated to different bins. For a triangle with sides $s, q, p$, the first contribution will go to a bin $(r_1, r_2)$ with $s$ playing the role of $r_1$ and $q$ that of $r_2$ (or vice versa; the estimate is constructed to be symmetric under this interchange). The second contribution will go to a bin $(r_1, r_2)$ with $q$ playing the role of $r_1$ and $p$ that of $r_2$, and third to a bin $(r_1, r_2)$ with $p$ playing the role of $r_1$ and $s$ that of $r_2$. In general, these three radial bin combinations could all be different. Hence, typically (save for equilateral triangles and isosceles triangles, each a set of measure zero), each triangle will contribute to a given radial bin with only $one$ of the three possible choices for line of sight. Hence, one will not have an opportunity for the cancellation discussed in \S\ref{sec:intro} and again in \S\ref{subsec:Scocc}, which occurs for the SPM estimator of the anisotropic 2PCF \citep{SE_wa_FT}.

\subsection{Sugiyama}
\cite{Sugiyama_2018} develops a basis similar to that in \cite{SE_aniso_3PCF}, but rather than tracking mixed harmonic coefficients with total angular momenta $\ell$, $\ell'$ and spins $m$, they track simply three angular momenta, $\ell$, $\ell'$ and $L$, where $L$ is given by the vector sum of the former two. $L$ is a measure of the anisotropy induced by RSD and \cite{Sugiyama_2018} argue that the main modes will be $L=0$ (isotropic bispectrum or 3PCF) and $L=2$ and $L=4$, as the Kaiser-formula RSD \citep{Kaiser1987} in the density are quadrupolar, and therefore will generate up to $L=4$ when one multiplies two density fields (the third density point in the triplet defines the local origin of coordinates and so does not count). $L$ corresponds to the angular momentum of a spherical harmonic in the line of sight, so their basis shows explicitly how the line of sight enters, in contrast to that of \cite{SE_aniso_3PCF}, where the line of sight enters by defining the rotation needed to place it along the $z$-axis at each successive primary.

As \cite{Sugiyama_2018} shows, the measurement of the anisotropic 3PCF in their basis can be recovered by a summing the coefficients as measured in the \cite{SE_aniso_3PCF} basis against a Wigner 3-$j$ symbol, and similarly, the \cite{SE_aniso_3PCF} coefficients that would be measured can be extracted from their basis by inversion of this sum using orthogonality of the 3-$j$ symbols. In this work, we build on the \cite{Sugiyama_2018} basis because we find the explicit appearance of the line of sight in it useful to develop our spherical harmonic expansion.

\iffalse
{\bf ZS: write!}

*Add in symmetrization point this is important; also mention it in the introduction. It applies to Roman's 2015 estimator too, by the way.
*Discuss degrees of freedom in context of Scoccimarro.

{\bf KG: write some text on the above point, which i s actually general and doesn't have to do with the basis chosen, really, but rather simply with the choice of los.}

\kg{Scocciomaro 2015's method, for example, rotates the line of sight among different members of a triplet, which would at first glance solve the symmetry problem. However, when instead of the bispectrum we are interested in computing the 3PCF (----isn't this a problem also for bispectrum? I remmember we discussed it but don't remmember why----), this symmetry is broken in the binning process. For example, if we first point the line of sight to the furthest galaxy in a determined triplet, the correlation between them will be added to a large scale bin. When the line of sight is shifted to another member of this triplet, its contribution will be added to another, shorter scale bin. In the end, the total result may be the same, but each contribution will not be allocated correctly.}
\fi

\section{Our Basis}
\label{sec:basis}

\begin{figure}
    \centering
    \includegraphics[width=8.5cm]{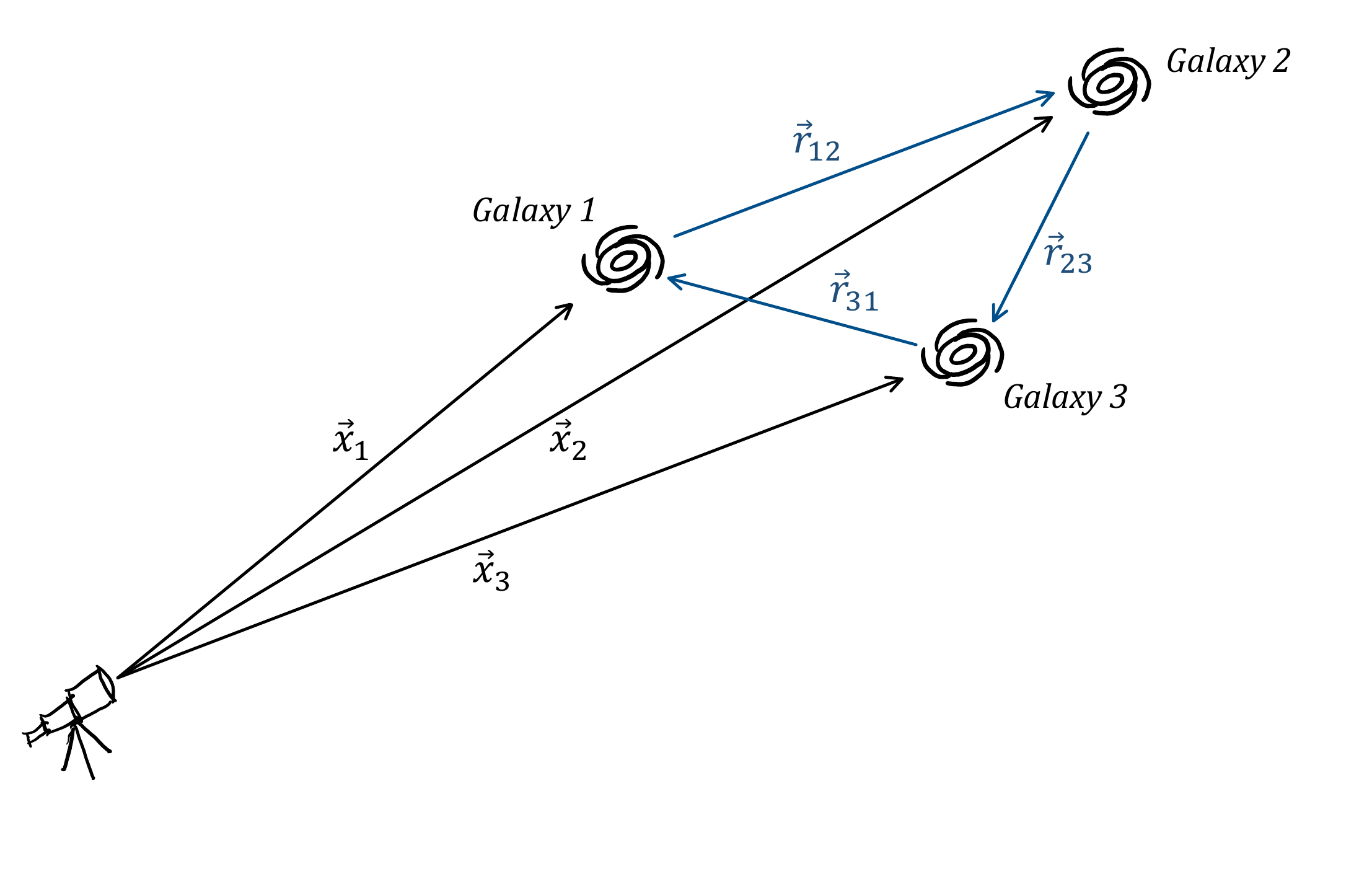}
    \caption{Diagram of a triplet of galaxies, each indicated by a line-drawing of a spiral. $\vec{x}_i$ is the position of the $i^{\rm th}$ galaxy with respect to the observer, who is indicated by the telescope at the lower left. The vectors $\vec{r}_{ij}$, shown in blue, are the relative separations between the three galaxies, as defined by equation (\ref{eqn:r's_definition}), where the order of the subscripts indicates the order of the subtraction, i.e. $\vec{r}_{12} = \vec{x}_2 - \vec{x}_1$, and $\vec{r}_{31} = \vec{x}_1 - \vec{x}_3$.}
    \label{fig:setup}
\end{figure}

Our estimate $\hat{\zeta}$ of the full 3PCF about a point $\vec{x}_1$ is
\begin{align}
\label{eqn:3pcf_est}
&\hat{\zeta}(r_{12}, r_{13}; \hat{r}_{12}, \hat{r}_{13}; \vec{x}_1) = \delta(\vec{x}_1) \delta(\vec{x}_1 + \vec{r}_{12})\delta(\vec{x}_1 + \vec{r}_{13}).
\end{align}
We note that this estimate has nine degrees of freedom on each side, so it has the full information on the three galaxies' positions. It is prior to performing any averaging either over rotations about the line of sight or over translations. $r_{12}$ and $r_{13}$ are the lengths of the two vectors extending from the density point at $\vec{x}_1$ to, respectively, the density points at $\vec{x}_2$ and $\vec{x}_3$. In general our convention is that in $\vr_{ij}$, the position vector corresponding to the first subscript is always subtracted from that corresponding to the second subscript.The geometry is shown in Figure \ref{fig:setup}. Explicitly, the relative position vectors about $\vec{x}_1$ are
\begin{align}
\label{eqn:r's_definition}
\vec{r}_{12} \equiv \vec{x}_2 - \vec{x}_1,\nonumber\\
\vec{r}_{13} \equiv \vec{x}_3 - \vec{x}_1.
\end{align}

\iffalse
We want to bin in radial bins, so we define the binned density field as 
\begin{align}
    \bar{\delta}(R; \hat{r}; \vec{x}) = \int r^2 dr \; \delta(\vec{x} + \vec{r}) \; \Phi(r; R)    
\end{align}
$\Phi(r; R)$ is a binning function that takes any density point at a distance $|\vec{r}|$ from $\vec{x}$ such that $|\vec{r}|$ is within the bin designated by $R$. For instance, $\Phi$ could be a spherical tophat that is non-zero if $R - \Delta R < r < R + \Delta R$, where $\Delta R$ is half the bin width, and zero otherwise. We may normalize $\Phi$ by dividing by the volume of the bin, $(4\pi/3)\left[(R + \Delta R)^3 -(R - \Delta R)^3\right]$. The dimensions of $\Phi$ will be inverse volume.
\fi

Our basis for the full 3PCF after averaging over rotations about the line of sight, which we denote $\hat{n}$, and over translations is, following \cite{Sugiyama_2018}
\begin{align}
    &\zeta(r_{12}, r_{13}, \hat{r}_{12}, \hat{r}_{13}, \hat{n}) = \sum_{\ell \ell_2 \ell_3} \zeta_{\ell \ell_2 \ell_3}(r_{12},r_{13}) \nonumber\\
    &\times \sum_{m_2 m_3} \begin{pmatrix}
\ell & \ell_2 & \ell_3\\
0 & m & -m
\end{pmatrix}
 Y_{\ell}^0(\hat{n}) Y_{\ell_2}^{m}(\hat{r}_{12})Y_{\ell_3}^{-m}(\hat{r}_{13}).
\end{align}
As already noted, $r_{12}$ and $r_{13}$ are ``internal'' parameters that describe the triangle side lengths. The $2 \times 3$ matrix is a Wigner 3-$j$ symbol, and the $Y_{\ell}^m$ are spherical harmonics, with our normalization and phase convention indicated in Appendix \ref{app:app1}. $\hat{n}$ is a unit vector giving the line of sight.

As discussed in \cite{Sugiyama_2018}, this basis captures all functions that have symmetry under rotation about the line of sight; that is why the spin associated with the angular momentum related to the spherical harmonic in $\hat{n}$ is zero in both the 3-$j$ symbol and the spherical harmonic. This point combined with the selection rule on the 3-$j$ symbol that the spins sum to zero is why the spherical harmonics involving $\ell_1$ and $\ell_2$ have equal and opposite spins. 

We now briefly explain how this is indeed a basis. In particular, one might wonder if one can truly extract the coefficient $\zeta_{\ell \ell_1 \ell_2}$ above using orthogonality to integrate each side against conjugated spherical harmonics of $\hat{n}$, $\hat{r}_{12}$ and $\hat{r}_{13}$. In particular, how this integration interplays with translation-averaging becomes less intuitive once one uses all three galaxy positions to define the line of sight, as we seek to do here.

Let us first see how this works in the case where only a single triplet member defines the line of sight, i.e. $\hat{n}\to \hat{x}_1$ for instance. Here, we have our local estimate of the 3PCF coefficient about $\vx_1$ as 
\begin{align}
\label{eqn:local_est_with_delta}
&\hat{\zeta}_{\ell \ell_2 \ell_3}(r_{12}, r_{23}; \vx_1) =\int d\Omega_1\; Y_{\ell}^{0*}(\hat{x}_1)\\
&\times \int d\Omega_{12} \;d\Omega_{13} \; Y^{m_2*}_{\ell_2}(\hat{r}_{12}) Y^{m_3*}_{\ell_3}(\hat{r}_{13})\;\delta(\vec{x}_1) \delta(\vec{x}_1 + \vr_{12}) \delta(\vec{x}_1 + \vr_{13}).\nonumber
\end{align}
We note that the lefthand side is in fact not a function of the full $\vx_1$ anymore (or, a trivial one), but only truly depends on $x_1$. The dependence on $\hat{x}_1$ has been projected onto $Y_{\ell}|^{0*}(\hat{x}_1)$. We may now translation-average as
\begin{align}
&\zeta_{\ell \ell_2 \ell_3}(r_{12}, r_{23}) = \frac{1}{V} \int x_1^2 dx_1\; \hat{\zeta}_{\ell \ell_2 \ell_3}(r_{12}, r_{23}; \vx_1),
\label{eqn:final_avg}
\end{align}
with $V$ the survey volume. 

Physically, this ordering of operations corresponded to sitting on a fixed global shell some distance $x_1$ away from the observer and then specializing to a galaxy at $\hat{x}_1$ on that shell. We then look for all pairs of galaxies distances $r_{12}, r_{13}$ away and project the angular structure around $\vec{x}_1$ onto harmonics. We then further co-add together all galaxy triplets whose primary was on the shell at $x_1$, weighting by a harmonic of $\hat{x}_1$. This is our ``local'' estimate $\hat{\zeta}_{\ell \ell_2 \ell_3}$ of the anisotropic 3PCF contribution from all triplets whose ``primary'' is a distance $x_1$ from the observer. The final averaging as in equation (\ref{eqn:final_avg}) is then over all global shells. Thus, we see that in this approach, the translation averaging really is done on global spherical shells about the observer. On each shell of fixed $x_1$, we average over rotations of $\hat{x}_1$ covering that whole shell.

Combining equations (\ref{eqn:local_est_with_delta}) and (\ref{eqn:final_avg}), we have
\begin{align}
\label{eqn:full_estimate}
&\zeta_{\ell \ell_2 \ell_3}(r_{12}, r_{23}) =\frac{1}{V} \int x_1^2 dx_1 \int d\Omega_1\; Y_{\ell}^{0*}(\hat{x}_1)\\
&\times \int d\Omega_{12} \;d\Omega_{13} \; Y^{m_2*}_{\ell_2}(\hat{r}_{12}) Y^{m_3*}_{\ell_3}(\hat{r}_{13})\;\delta(\vec{x}_1) \delta(\vec{x}_1 + \vr_{12}) \delta(\vec{x}_1 + \vr_{13}).\nonumber
\end{align}
Consider a change of variables from $\vec{x}_1$ to $\vec{n}$, where $\hat{n}$ is the direction vector not to a given galaxy within a triplet, but to some point within the plane of the triangle formed by the triplet. 

Now, to see that the directions $\hat{n}$ and $\hat{r}_{12}$, $\hat{r}_{13}$ are independent, which is what we need to extract our expansion coefficients, consider the following. At fixed $\hat{n}$, say, pointing to the centroid of the galaxy triplet, one is free to rotate $\hat{r}_{12}$ and $\hat{r}_{13}$ at will. This will pull their common origin point, $\vec{x}_1$, along with them, but that is fine; we do not need it. One should imagine a triangular slice of cheese pierced by a toothpick at its center. One can hold the toothpick pointing to any direction one likes, and independently rotate the cheese slice on it so that the two vectors, defining two of its sides from a given vertex, rotate freely. The vertex from which they stem will also rotate, but that is fine. We need only that $\hat{n}$ is independent from $\hat{r}_{12}$ and $\hat{r}_{13}$ to invoke orthogonality of the spherical harmonics.

Hence, formally, at a given location in space $\vec{n}$, we are free to look for all galaxy triplets with that centroid, and accumulate them onto bins in $r_{12}$ and $r_{13}$ with weights given by spherical harmonics of $\hat{n}$, $\hat{r}_{12}$, and $\hat{r}_{13}$.

\iffalse
{\bf ZS: need to add discussion of basis-hood, i.e. that we can integrate independently over $\hat{n}$, and the $\hat{r}_{ij}$.}
\fi
\iffalse
Our basis for the 3PCF is, around $\vec{x}$, to expand it as 
\begin{align}
&\hat{\zeta}(\vec{r}_{12}, \vec{r}_{13}; \vec{x}) =   \sum_{\ell_1 \ell_2 L} \hat{\zeta}_{\ell_1 \ell_2 L} (r_{12}, r_{13}; \vec{x}) \sum_{m_1 m_2 m} 3j-symbol\;of\;\ell\;and\;m\;\nonumber\\
&\times Y_{\ell_1}^{m_1}(\hat{r}_{12})
Y_{\ell_2}^{m_2}(\hat{r}_{13}) Y_L^M(\hat{n}) . 
\end{align}
$\hat{n}$ is the line of sight to the triangle; we discuss precisely how to define it shortly. This is the Sugiyama basis (their equation 4, but without the normalization factors they include in each spherical harmonic to find $y_{\ell}^{m}$ rather than $Y_{\ell}^{m}$, and translated to configuration space rather than Fourier space).

We may bin both sides of the above radially without affecting the angular structure---actually, we may not! because our expansion for $Y_L^M(\hat{r})$ includes the lengths of $r_{12}$ and $r_{13}$, so they will get integrated over when we bin!!
\fi

\section{Choice of Line of Sight}
\label{sec:los}

%{\bf ZS. Motivate by referencing plot in Sugiyama appendix.}
We may treat two cases for $\hat{n}$ that go beyond taking $\hat{n}$ as $\hat{x}$ as was done in \cite{Scoccimarro_2015}, \cite{SE_aniso_3PCF}, and \cite{Sugiyama_2018}. We may look at the generalization of the angle bisector definition of the line of sight  for the anisotropic 2PCF, which here reads
\begin{align}
\hat{n}_{U} = \frac{\hat{x}_1 + \hat{x}_1 + \hat{x}_2}{|\hat{x}_1 + \hat{x}_1 + \hat{x}_2|} ,
\end{align}
where subscript $U$ in $\hat{n}_U$ denotes the ``unit-vector-average'' method. We may also look at the analog of the midpoint definition of the line of sight for the anisotropic 2PCF: 
\begin{align}
\label{eqn:main_nc}
\hat{n}_C = \frac{\vec{x}_1 + \vec{x}_1 + \vec{x}_2}{|\vec{x}_1 + \vec{x}_1 + \vec{x}_2|} .
\end{align}
In this case, subscript $C$ in $\hat{n}_C$ denotes the ``centroid'' method; as we will prove, this line of sight intersects the triplet of galaxies at its centroid.

\subsection{Proof that Position-Vector-Average Leads to the Centroid}
\begin{figure}
    \centering
    \includegraphics[width=8.5cm]{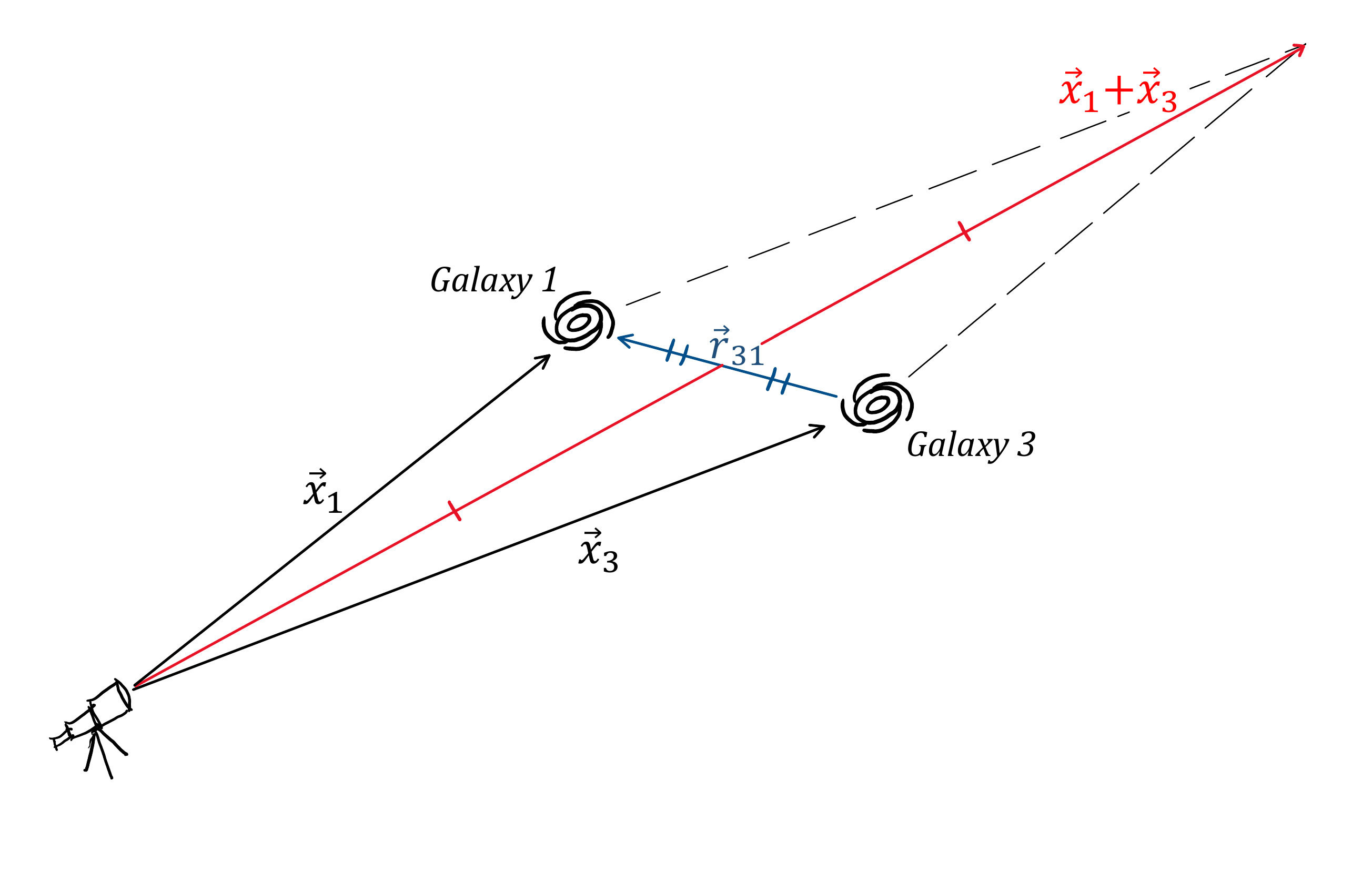}
    \caption{Diagram showing that the modulus of the sum of two galaxy positions $\vec{x}_1$ and $\vec{x}_3$ will be one of the diagonals of a parallelogram with sides $x_1$ and $x_3$. It is a theorem that each diagonal of a parallelogram bisects the other, so the sum $\vec{x}_1 + \vec{x}_3$ will bisect $\vr_{31}$.}
    \label{fig:parallelogram}
\end{figure}

\begin{figure*}
    %\centering
    \includegraphics[width=17.5cm]{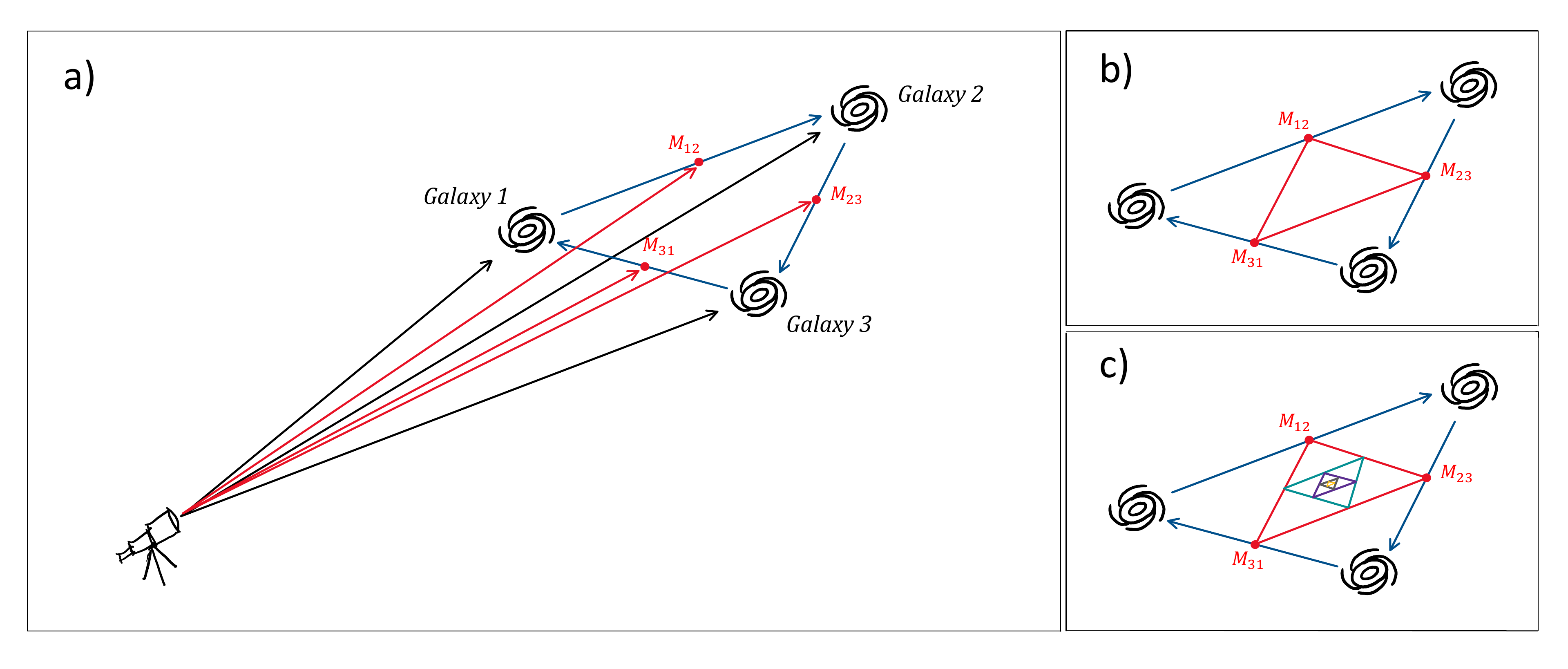}
    \caption{Illustration of the proof that a line of sight taken as the sum of the three galaxy position vectors, as in equation (\ref{eqn:main_nc}), will intersect the centroid of the base triangle formed by the three galaxies. In panel (a) we are summing each pair of positions as $\vec{x}_i + \vec{x}_j $, and the vector given by half this sum intersects the base triangle at the points indicated in red and labeled $M_{ij}$. These points are in fact the midpoints of the three base triangle sides, due to the fact that each sum $\vec{x}_i + \vec{x}_j$ could be represented  by a parallelogram (as shown in Figure \ref{fig:parallelogram}), whose diagonals always bisect each other. In panel (b) we connect these three intersection points, to form a triangle within the base triangle. It turns out that, since the $M_{ij}$ are midpoints, this $inner$ triangle is known as the medial triangle. The medial triangle of a given parent triangle will always share the same centroid. In panel (c) we repeat the process shown in  (a) and (b) but now starting with the medial triangle as our new mother triangle. Panel (c) shows a number of iterations of this process. We can develop smaller and smaller concentric medial triangles by forming more and more sums of pairwise sums of the original vectors $\vec{x}_i$ and $\vec{x}_j$. These sums will all be rescalings of our desired line of sight, and hence pass through the base at the same point. Ultimately this point will be the only point remaining enclosed as we go to arbitrarily small medial triangles, meaning that our line of sight must pass through the centroid.}
    \label{fig:x1x2x3_medians}
\end{figure*}

First, we consider summing the pairwise sums of the galaxies' position vectors: $(\vec{x}_1+\vec{x}_2)$, $(\vec{x}_2+\vec{x}_3)$, $(\vec{x}_3+\vec{x}_1)$. If we then sum up the resulting vectors, we will obtain a vector that has the same direction as $\vx_1$, $\vx_2$ and $\vx_3$, i.e.
\begin{align}
    (\vec{x}_1+\vec{x}_2) + (\vec{x}_2+\vec{x}_3) + (\vec{x}_3+\vec{x}_1) = 2(\vec{x}_1+\vec{x}_2+\vec{x}_3),
\end{align}
and
\begin{align}
    \reallywidehat{2(\vec{x}_1+\vec{x}_2+\vec{x}_3)} = \reallywidehat{\vec{x}_1+\vec{x}_2+\vec{x}_3} .
\end{align}
As one can see in Figure \ref{fig:parallelogram}, parallelograms' diagonals bisect each other, so these vectors will split each side of the base triangle (i.e. the triangle of galaxies) in two. Put another way, they will intersect each side of the base triangle at its midpoint.  If we then connect these midpoints, we form a medial triangle. It is a theorem that the medial triangle has the same centroid as its parent triangle.\footnote{For discussion of the medial triangle's properties, see \url{https://mathworld.wolfram.com/MedialTriangle.html}} We can repeat this process within the medial triangle. We consider the sum $[(\vx_1 + \vx_2) + (\vx_2 + \vx_3)]$$+ [(\vx_2 + \vx_3) + (\vx_3 + \vx_1)]$$+ [(\vx_3 + \vx_1) + (\vx_1 + \vx_2)]$, which will have the same direction as our original line of sight. Applying the same logic as with the first medial triangle, we see that each pair in brackets will intersect the medial triangle's sides at their midpoints. We can then use these midpoints to define a new, smaller medial triangle. By performing successive sums of pairs of pairs of pairs $ad$ $infinitum$, we can construct smaller and smaller medial triangles by successive bisections. Each will share the same centroid as the parent triangle of three galaxies. In the infinite-iteration limit, the medial triangle so constructed will be arbitrarily small and yet our line of sight must pass through it. The centroid of the parent triangle of galaxies must also. Hence, we have shown that our line of sight is arbitrarily close to the centroid: thus, it must pass through the centroid. We show this proof visually in Figure \ref{fig:x1x2x3_medians}.

%\begin{figure*}
%\centering
%    \includegraphics[width=0.8\linewidth]{figures/x1x2x3_medians.jpg}\hfil
%    \includegraphics[width=0.47\linewidth]{figures/x1x2x3_medians_2.jpg}\hfil
%    \includegraphics[width=0.47\linewidth]{figures/x1x2x3_medians_3.jpg}\par\medskip
%\caption{}
%\label{fig:imgs2}
%\end{figure*}

\subsection{Centroid vs. Unit-Vector-Average Methods}
We now seek to determine at what order our two methods differ from each other. We compute find $\hat{n}_C - \hat{n}_U$, which is
\begin{align}
    \hat{n}_C-\hat{n}_U &= \frac{\vec{n}_{\rm C}}{|\vec{n}_{\rm C}|} - \frac{\vec{n}_{\rm U}}{|\vec{n}_{\rm U}|} \\
    &= \frac{\vec{x}_1+\vec{x}_2+\vec{x}_3}{|\vec{x}_1+\vec{x}_2+\vec{x}_3|} - \frac{\hat{x}_1+\hat{x}_2+\hat{x}_3}{|\hat{x}_1+\hat{x}_2+\hat{x}_3|} .
    \nonumber 
\end{align}

We first examine $\hat{n}_C$. Substituting $\vec{x}_2 = \vec{x}_1+\vec{r}_{12}$ and $\vec{x}_3 = \vec{x}_1+\vec{r}_{13}$ based on equation (\ref{eqn:r's_definition}), and simplifying, we find
\begin{align}
    \label{eqn:nc}
    \vec{n}_{\rm C} &= 3\vec{x}_1+\vec{r}_{12}+\vec{r}_{13}\nonumber \\
    &= 3x_1[\hat{x}_1+\vec{r}_{12}/(3x_1)+\vec{r}_{13}/(3x_1)] .
\end{align}
Defining
\begin{align}
    &\epsilon_{12,1} \equiv r_{12}/(3x_1) , \nonumber \\
    &\epsilon_{13,1} \equiv r_{13}/(3x_1) ,
\end{align}
we can turn equation (\ref{eqn:nc}) into
\begin{align}
        \vec{n}_{\rm C} &= \hat{x}_1+\epsilon_{12,1}\hat{r}_{12}+\epsilon_{13,1}\hat{r}_{13} .
\end{align}
Since we want $\hat{n}_C = \vec{n}_{\rm C}/|\vec{n}_{\rm C}|$, we will need to expand the denominator of this equation:
\begin{align}
    |\vec{n_C}|^{-1} &= (|\hat{x}_1+\epsilon_{12,1}\hat{r}_{12}+\epsilon_{13,1}\hat{r}_{13}|^2)^{-1/2} \\
    &= [1 + 2\hat{x}_1\cdot(\epsilon_{12,1}\hat{r}_{12}+\epsilon_{13,1}\hat{r}_{13}) + \mathcal{O}(\epsilon^2)]^{-1/2} \nonumber \\
    &\approx 1 - \epsilon_{12,1}\hat{r}_{12}\cdot\hat{x}_1 - \epsilon_{13,1}\hat{r}_{13}\cdot\hat{x}_1 \nonumber ,
\end{align}
to finally obtain
\begin{align}
    &\hat{n}_C = (\hat{x}_1+\epsilon_{12,1}\hat{r}_{12}+\epsilon_{13,1}\hat{r}_{13})(1 - \epsilon_{12,1}\hat{r}_{12}\cdot\hat{x}_1 - \epsilon_{13,1}\hat{r}_{13}\cdot\hat{x}_1) \nonumber \\
    &= \hat{x}_1 + \epsilon_{12,1}[\hat{r}_{12}-(\hat{r}_{12}\cdot\hat{x}_1)\hat{x}_1] + \epsilon_{13,1}[\hat{r}_{13}-(\hat{r}_{13}\cdot\hat{x}_1)\hat{x}_1] .
\end{align}

We now examine $\hat{n}_U$. We can rewrite the direction vector to the galaxy at $\vec{x}_2$ as 
\begin{align}
    \hat{x}_2 &= \frac{\vec{x}_1+\vec{r}_{12}}{|\vec{x}_1+\vec{r}_{12}|} 
    = \frac{\hat{x}_1+3\epsilon_{12,1}\hat{r}_{12}}{|\hat{x}_1+3\epsilon_{12,1}\hat{r}_{12}|} ,
\end{align}
and at $\vec{x}_3$ as
\begin{align}
    \hat{x}_3 &= \frac{\vec{x}_1+\vec{r}_{13}}{|\vec{x}_1+\vec{r}_{13}|} 
    = \frac{\hat{x}_1+3\epsilon_{13,1}\hat{r}_{13}}{|\hat{x}_1+3\epsilon_{13,1}\hat{r}_{13}|} .
\end{align}
Thus $\vec{n}_{\rm U}$ will be
\begin{align}
    \vec{n}_{\rm U} &= \hat{x}_1+\frac{\hat{x}_1+3\epsilon_{12,1}\hat{r}_{12}}{|\hat{x}_1+3\epsilon_{12,1}\hat{r}_{12}|}+\frac{\hat{x}_1+3\epsilon_{13,1}\hat{r}_{13}}{|\hat{x}_1+3\epsilon_{13,1}\hat{r}_{13}|} .
\end{align}
Expanding $|\hat{x}_1+3\epsilon_{21,2}\hat{r}_{12}|^{-1}$, we find
\begin{align}
    |\hat{x}_1+3\epsilon_{21,2}\hat{r}_{12}|^{-1} &= [(\hat{x}_1+3\epsilon_{12,1}\hat{r}_{12})^2]^{-1/2} \nonumber \\
    &= [1 + 6 \epsilon_{12,1}\hat{r}_{12} \cdot \hat{x}_1 + (3\epsilon_{12,1})^2]^{-1/2} \nonumber \\
    &\approx 1 - 3\epsilon_{12,1}\hat{r}_{12} \cdot \hat{x}_{1} + \mathcal{O}\left(\epsilon_{12,1}^2\right) ,
\end{align}
which leads to
\begin{align}
    \vec{n}_{\rm U} &= \hat{x}_1+(\hat{x}_1+3\epsilon_{12,1}\hat{r}_{12})(1-3\epsilon_{12,1}\hat{r}_{12}\cdot\hat{x}_{1}) \nonumber\\
    &+(\hat{x}_1 +3\epsilon_{13,1}\hat{r}_{13})(1-3\epsilon_{13,1}\hat{r}_{13}\cdot\hat{x}_1) \nonumber \\
    &= 3\{\hat{x}_1+\epsilon_{12,1}[\hat{r}_{12}-\hat{x}_1 (\hat{r}_{12}\cdot \hat{x}_1)] \nonumber \\
    &+\epsilon_{13,1}[\hat{r}_{13}-\hat{x}_1 (\hat{r}_{13}\cdot \hat{x}_1)]\}
\end{align}
In order to find $\hat{n}_U = \vec{n}_{\rm U}/|\vec{n}_{\rm U}|$, we also need to expand the denominator of this equation as in
\begin{align}
    |\vec{n}_{\rm U}|^{-1} &= \frac{1}{3}\bigg\{\bigg[\hat{x}_1+\epsilon_{12,1}[\hat{r}_{12}-\hat{x}_1 (\hat{r}_{12}\cdot \hat{x}_1)] \nonumber \\
    &+\epsilon_{13,1}[\hat{r}_{13}-\hat{x}_1 (\hat{r}_{13}\cdot \hat{x}_1)]\bigg]^2\bigg\}^{-1/2} \\
    &= \frac{1}{3} \{ 1+\frac{2}{3} [3\epsilon_{12,1}(\hat{r}_{12}\cdot\hat{x}_1-\hat{r}_{12}\cdot\hat{x}_1)(\hat{x}_1 \nonumber \\
    &- \epsilon_{12,1}[\hat{r}_{12}-\hat{x}_1(\hat{r}_{12}\cdot \hat{x}_1)]-\epsilon_{13,1} [\hat{r}_{13}-\hat{x}_1(\hat{r}_{13}\cdot \hat{x}_1)])^2]\} .
    \nonumber
\end{align}
This has the form $\hat{x}_1^2+2\hat{x}_1\cdot\vec{v}+\vec{v}^2$, where $\vec{v}$ is here just a general vector. At $\mathcal{O}(\epsilon)$, we have $|\vec{n}_{\rm U}|^{-1} = 1$. So up to (inclusive) $\mathcal{O}(\epsilon)$, we will have
\begin{align}
    \hat{n}_U \approx \hat{x}_1 &+ \epsilon_{12,1}[\hat{r}_{12}-\hat{x}_1(\hat{r}_{12}\cdot \hat{x}_1)] \\
    &+\epsilon_{13,1}[\hat{r}_{13} -\hat{x}_1(\hat{r}_{13}\cdot \hat{x}_1)] ,
    \nonumber
\end{align}
which is exactly what we found for $\hat{n}_C$. This means that
\begin{align}
    \hat{n}_C - \hat{n}_U = 0
\end{align}
up to order $\mathcal{O}(\epsilon^2)$.

\begin{figure*}
    %\centering
    \includegraphics[width=18cm]{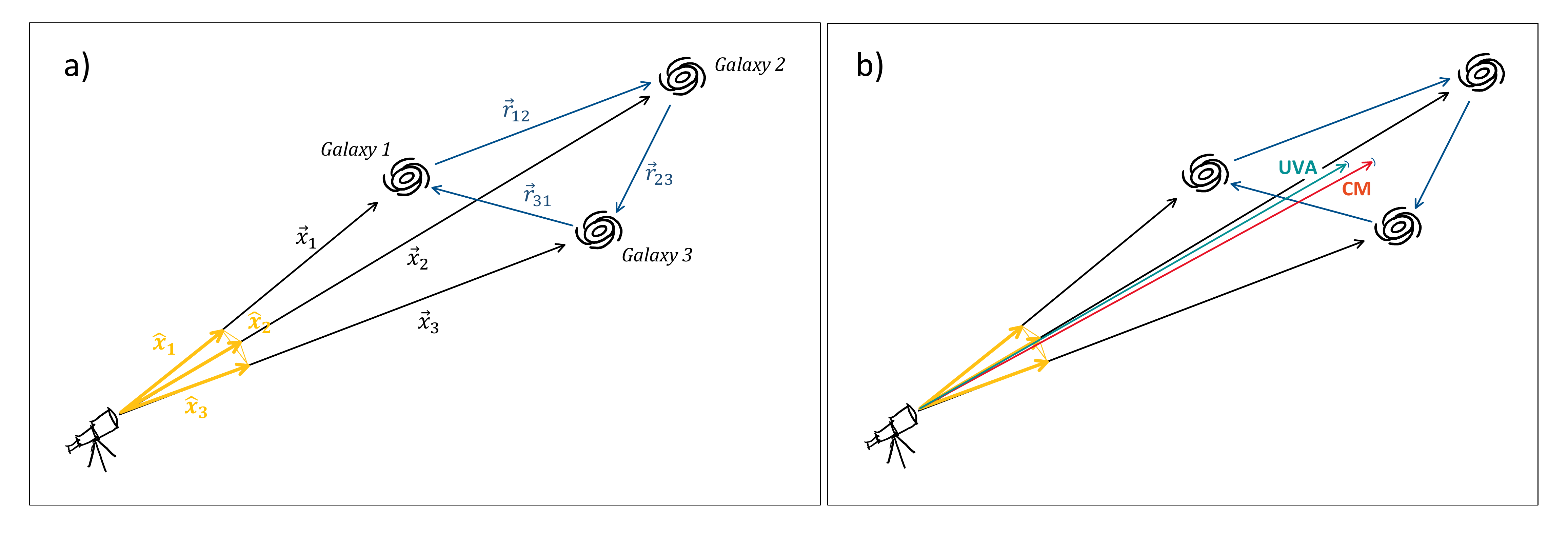}
    \caption{Representation of the unit vectors $\hat{x}_1$, $\hat{x}_2$, and $\hat{x}_3$ (yellow) in panel (a), and a comparison between the Centroid Method (CM) line-of-sight (which crosses the centroid of the triangle formed by the three galaxies; in teal), and the Unit-Vector-Average (UVA) method (scarlet) in panel (b).}
    \label{fig:cm_uva}
\end{figure*}

\section{Factorization of our basis}
\label{sec:factorization}
To extract the coefficient $\hat{\zeta}_{\ell_1 \ell_2 L}(r_{12}, r_{13}; \vec{x})$ we may use orthogonality:
\begin{align}
\label{eqn:full_extract}
&\zeta_{\ell \ell_2 \ell_3}(r_{12}, r_{23}) =\frac{1}{V} \int n^2 dn \int d\Omega_n\; Y_{\ell}^{0*}(\hat{n})\\
&\times \int d\Omega_{12} \;d\Omega_{13} \; Y^{m_2*}_{\ell_2}(\hat{r}_{12}) Y^{m_3*}_{\ell_3}(\hat{r}_{13})\;\delta(\vec{x}_1) \delta(\vec{x}_1 + \vr_{12}) \delta(\vec{x}_1 + \vr_{13}).\nonumber
\end{align}
where we now understand $\vec{x}_1$ to be a function of $\vec{n}$ and $\vec{r}_{12}$ and $\vec{r}_{13}$.

However , actually evaluating the above seems to require a triplet count: we have to look at the triplet of points $\vec{x}_1$, $\vec{x}_1 + \vec{r}_{12}$, and $\vec{x}_1 + \vec{r}_{13}$ to construct $\hat{n}$ if we set $\hat{n} = \hat{n}_{\rm C}$ or $\hat{n}_{\rm U}$. 

We ask if $\vec{n}$ can be rewritten as a function of $\vec{x}_1, \vec{r}_{12}$ and $\vec{r}_{13}$. We then ask if the spherical harmonic of $\hat{n}$ can be factorized into a product of functions of these, so that the integrand factorizes. If these were possible, we could then rewrite the integration as one over $d^3\vx_1\; d\Omega_{12} \; d\Omega_{13}$ and compute each integration separately. We could sit at a point $\vx_1$ perform the inner two integrals independently from each other, and hence obtain an algorithm scaling as $N^2$ much in the same way that \cite{SE_3pt_alg} does.
Hence, we rewrite the triple product of spherical harmonics on the righthand side above in terms of $\hat{x}_1$, $\hat{r}_{12}$, and $\hat{r}_{13}$. This may be done using the results in our Appendix \ref{app:app1}, of which we duplicate the key equations here.
\begin{align}
    \label{eqn:main_in_body}
    Y_\ell^m &(\reallywidehat{3\vec{x}_1 + \vec{r}_{12} + \vec{r}_{13}}) =  4\pi \sum_{n=0}^\infty \sum_{k=0}^{\lfloor n/2 \rfloor} \; \sum_{z=0}^{n-2k} \; \sum_{z'=0}^{k} \; \sum_{z''=0}^{k-z'} \; \sum_{j=z'\downarrow 2} \nonumber \\
    & \times \sum_{m=-j}^{j} \; \sum_{j'=(n-2k-z) \downarrow 2} \; \sum_{m'=-j'}^{j'} \; \sum_{j''=z \downarrow 2} \; \sum_{m''=-j''}^{j''} \; \sum_{\lambda=0}^{\ell} \; \sum_{\lambda'=0}^{\ell-\lambda} \nonumber \\
    & \times \sum_{\mu = - \lambda}^{\lambda} \; \sum_{\mu' = -\lambda'}^{\lambda'} \; \sum_{L_0 M_0} \; \sum_{\mathcal{L}_0 \mathcal{M}_0} \; \sum_{L_1 M_1} \; \sum_{\mathcal{L}_1 \mathcal{M}_1} \; \sum_{L_2 M_2} \; \sum_{\mathcal{L}_2 \mathcal{M}_2} \nonumber \\ \nonumber \\ \nonumber
    & \times \mathcal{U}_{z z' z'' j j' j''}^{k n \; (\ell/2)}  \; Q_{\ell \lambda \lambda'} \; \mathcal{G}_{\lambda \lambda' j j' j'' L_0 \mathcal{L}_0 L_1 \mathcal{L}_1 \ell L_2 \mathcal{L}_2}^{\mu \mu' m m'm'' M_0 \mathcal{M}_0 M_1 \mathcal{M}_1 M_2 \mathcal{M}_2} \nonumber \\ \nonumber \\
    & \times \epsilon_{12,1}^{n-z-z'-2z''+\lambda'} \; \epsilon_{13,1}^{z+z'+2z''+\ell-\lambda-\lambda'} \\ \nonumber \\ \nonumber
    & \times Y_{L_0}^{M_0}(\hat{x}_1) \; Y_{L_1}^{M_1}(\hat{r}_{12}) \;  Y_{L_2}^{M_2}(\hat{r}_{13}) , \\ \nonumber
\end{align}
where
\begin{align}
    \mathcal{U}_{z z' z'' j j' j''}^{k n \alpha} \equiv \frac{4\pi^2}{(2j'+1)(2j''+1)} \; \mathcal{T}_{z z' z''}^{j k n \alpha} \; \mathcal{S}_{j'}^{n-2k-z} \; \mathcal{S}_{j''}^{z} ,
\end{align}
\begin{align}
    &\mathcal{T}_{z z' z''}^{j k n \alpha} \equiv \frac{\Gamma(n-k+\alpha)}{\Gamma(\alpha)}\\
    &\times \frac{4\pi \; (-1)^k \; (2)^{z'+2n-4k-(z'-j)/2}}{z! \; z''! \; (n-2k-z)! \; (k-z'-z'')! \; [(z'-j)/2]! \; (j+z'+1)!!}
    \nonumber
\end{align}
\begin{align}
    \mathcal{S}_j^{z'} \equiv \frac{(2j+1)z'!}{2^{(z'-j)/2}((z'-j)/2)! (j+z'+1)!!}.
\end{align}
\begin{align}
    \mathcal{Q}_{\ell \lambda \lambda'} &\equiv     \begin{pmatrix}
    2\ell\\ 
    2\lambda
    \end{pmatrix}
    ^{1/2}
    \begin{pmatrix}
    2(\ell-\lambda)\\ 
    2\lambda'
    \end{pmatrix}
    ^{1/2} (-1)^{\lambda + \mu} \sqrt{2\ell + 1} \sqrt{2(\ell - \lambda) + 1} \nonumber\\
& \times  \begin{pmatrix}
\lambda & \ell - \lambda & \ell\\
\mu & m-\mu & -m
\end{pmatrix} \begin{pmatrix}
\lambda' & \ell - \lambda - \lambda' & \ell - \lambda\\
\mu' & m - \mu - \mu' & -m + \mu
\end{pmatrix},
\end{align} 
and
\begin{align}
    \label{eqn:angmom_in_body}
    \mathcal{G}&_{\lambda \lambda' j j' j'' L_0 \mathcal{L}_0 L_1 \mathcal{L}_1 \ell L_2 \mathcal{L}_2}^{\mu \mu' m m'm'' M_0 \mathcal{M}_0 M_1 \mathcal{M}_1 M_2 \mathcal{M}_2} \equiv \\
    & \hspace{2cm} \mathcal{G}_{\lambda j' \mathcal{L}_0}^{\mu -m' -\mathcal{M}_0} \; \mathcal{G}_{\mathcal{L}_0 j'' L_0}^{\mathcal{M}_0 -m'' -M_0} \; \mathcal{G}_{\lambda' j \mathcal{L}_1}^{\mu' m -\mathcal{M}_1} \nonumber \\
    & \hspace{2cm} \times \mathcal{G}_{\mathcal{L}_1 j' L_1}^{\mathcal{M}_1 m' -M_1} \; \mathcal{G}_{(\ell-\lambda-\lambda') j \mathcal{L}_2}^{(m-\mu-\mu') -m -\mathcal{M}_2} \; \mathcal{G}_{\mathcal{L}_2 j'' L_2}^{\mathcal{M}_2 m'' -M_2}. \nonumber \\ \nonumber
\end{align}
The summation indices of equation (\ref{eqn:main_in_body}), their limits, and detailed explanation can be found in Table \ref{table1}. We present here only the result for the Centroid method because the Unit-Vector-Average method agrees with it at $\mathcal{O}(\theta^2)$, as shown in \S\ref{sec:los}. However if one wished to go beyond that order, applying the math in Appendix \ref{app:app1} to do so is straightforward.

\section{Leading-Order Correction from Our Approach}
\label{sec:leading}
Since $\epsilon_{12,1}$ and $\epsilon_{13,1}$ are of the same order, we can suppose that they represent the same variable and analyze the sum of their powers to find the leading-order correction required by our approach, i.e. to find all terms that lead to a contribution at $\mathcal{O}(\epsilon)$. Taking it that $\epsilon_{12,1}\simeq\epsilon_{13,1}\simeq\epsilon$, we see that equation (\ref{eqn:main_in_body}) behaves as
\begin{align}
   \epsilon^{n-z-z'-2z''+\lambda'}\epsilon^{z+z'+2z''+\ell-\lambda-\lambda'} \simeq \epsilon^{n+\ell-\lambda} .
\end{align}
We see that at $\mathcal{O}(\epsilon)$, there are two possible cases that can contribute. We must have $n+\ell-\lambda=1$. The first way this can happen is when $n=0$ and $\ell-\lambda=1$. In this first case we reduce equation (\ref{eqn:main_in_body}) to
\begin{align}
\label{eqn:case_n1}
    Y_\ell^m &(\reallywidehat{3\vec{x}_1 + \vec{r}_{12} + \vec{r}_{13}}) =  4\pi \; \delta^{\rm K}_{n, 0} \;  \delta^{\rm K}_{k, 0} \; \delta^{\rm K}_{z, 0} \; \delta^{\rm K}_{z', 0} \; \delta^{\rm K}_{z'', 0} \nonumber \\
    &\times \delta^{\rm K}_{j, 0} \; \delta^{\rm K}_{m, 0} \; \delta^{\rm K}_{j’, 0} \; \delta^{\rm K}_{m', 0} \; \delta^{\rm K}_{j’', 0} \; \delta^{\rm K}_{m'', 0} \; \delta^{\rm K}_{\lambda, \ell-1} \\
    &\times \sum_{\lambda'=0}^{1} \; \sum_{\mu = 1-\ell}^{\ell-1} \; \sum_{\mu' = \lambda'}^{\lambda'} \; \sum_{L_0 M_0} \; \sum_{\mathcal{L}_0 \mathcal{M}_0} \; \sum_{L_1 M_1} \; \sum_{\mathcal{L}_1 \mathcal{M}_1} \; \sum_{L_2 M_2} \; \sum_{\mathcal{L}_2 \mathcal{M}_2} \nonumber \\ \nonumber \\
    & \times \mathcal{C} \; \epsilon \; Y_{L_0}^{M_0}(\hat{x}_1) \; Y_{L_1}^{M_1}(\hat{r}_{12}) \;  Y_{L_2}^{M_2}(\hat{r}_{13}) , \nonumber
\end{align}
$\delta^{\rm K}_{i,j}$ is a Kronecker delta, unity when its subscripts are equal and zero otherwise. We use the Kronecker delta to explicitly indicate all of the sum indices that get set by forcing $n = 0$. $\mathcal{C}$ is a placeholder for the complicated product of defined coefficients coming from equation (\ref{eqn:main_in_body}); we give explicit results for it shortly. We notice that in the above, there are two possibilities for $\lambda'$; we consider both. The sums over the angular momenta and spins are all governed by the angular momentum coupling represented by the product of six Gaunt integrals encoded in equation (\ref{eqn:angmom_in_body}). For short, we denote this equation by $\mathcal{G}_{[12]}$ in what follows, where the subscript $''[12]''$ is because it has 12 subscripts (and we wish to distinguish it clearly from a single Gaunt integral). The angular momentum couplings ultimately set which spherical harmonic coefficients of the density field, given by momenta and spins $L_0 M_0, L_1 M_1$, and $L_2 M_2$ we will need to compute.

We now look at the reduction of equation (\ref{eqn:main_in_body}) in our second case, when $n=1$ and $\ell-\lambda=0$:
\begin{align}
\label{eqn:case_n0}
    Y_\ell^m &(\reallywidehat{3\vec{x}_1 + \vec{r}_{12} + \vec{r}_{13}}) =  4\pi \; \delta^{\rm K}_{n, 1} \;  \delta^{\rm K}_{k, 0} \;  \delta^{\rm K}_{z', 0} \; \delta^{\rm K}_{z'', 0} \nonumber \\
    &\times \delta^{\rm K}_{j, 0} \; \delta^{\rm K}_{m, 0} \delta^{\rm K}_{\lambda', 0} \; \sum_{z=0}^{1} \;  \sum_{m'=-j'}^{j'} \; \delta^{\rm K}_{j’, 1-z} \; \delta^{\rm K}_{j’', z} \; \sum_{m''=-j''}^{j''} \; \sum_{\mu = - \ell}^{\ell} \nonumber \\
    &\times \sum_{L_0 M_0} \; \sum_{\mathcal{L}_0 \mathcal{M}_0} \; \sum_{L_1 M_1} \; \sum_{\mathcal{L}_1 \mathcal{M}_1} \; \sum_{L_2 M_2} \; \sum_{\mathcal{L}_2 \mathcal{M}_2} \\
    & \times \mathcal{C} \; \epsilon \; Y_{L_0}^{M_0}(\hat{x}_1) \; Y_{L_1}^{M_1}(\hat{r}_{12}) \;  Y_{L_2}^{M_2}(\hat{r}_{13}). \nonumber
\end{align}
Again we have used Kronecker deltas to explicitly all of the summation indices whose values are determined by setting $n=1$. We see that here, $\lambda'$ is fixed but there are now two values of $z$, $0$ and $1$, that we must consider.

\begin{figure}
    \centering
    \includegraphics[width=8.5cm]{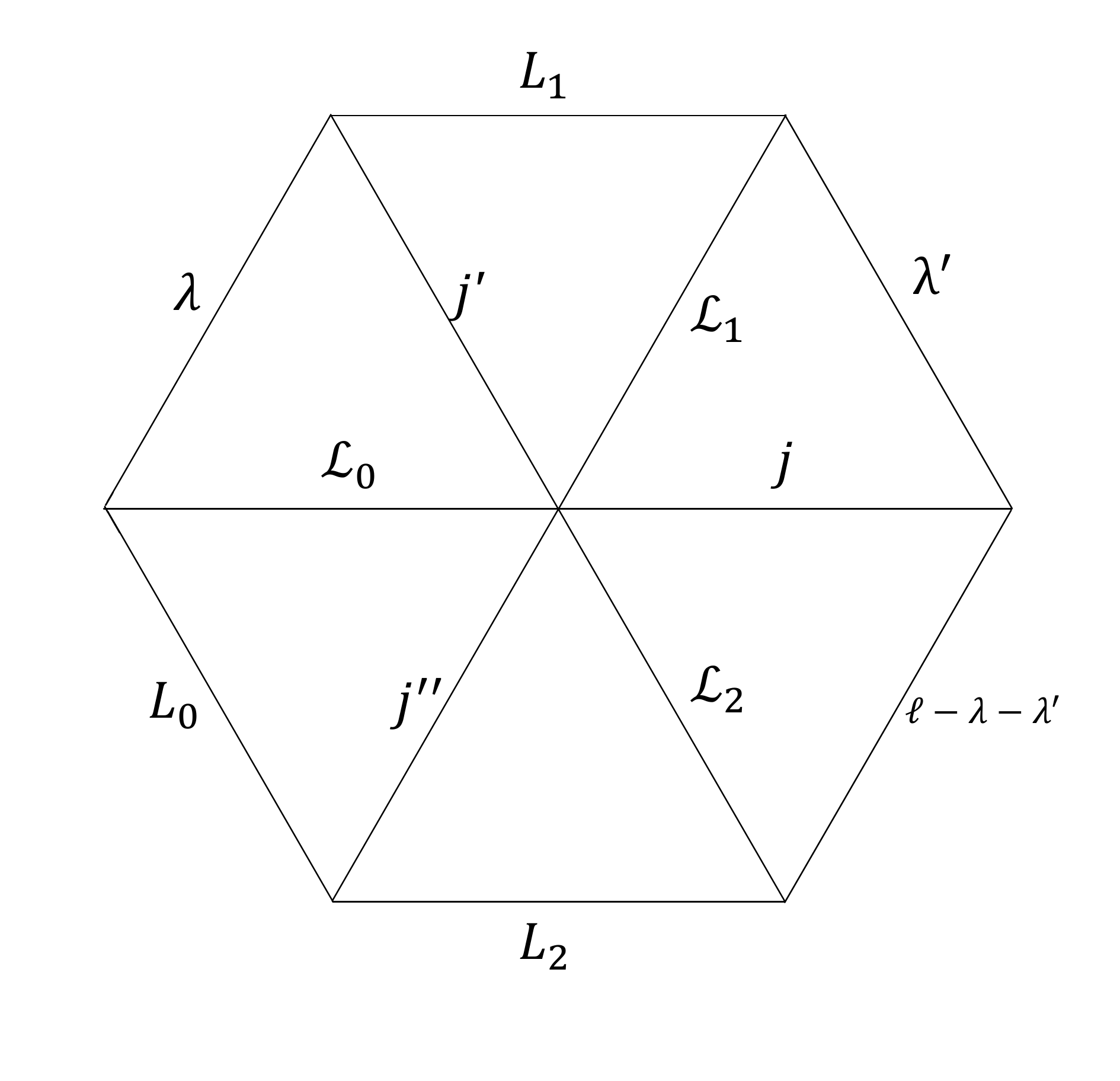}
    \caption{Diagram showing  the angular momentum coupling structure indicated by equation (\ref{eqn:angmom_in_body}). We note that though it looks similar, this is not a Yutsis diagram (e.g. \citealt{Yutsis_1962}; see also \citealt{Cahn_2020}). We have displayed each 3-$j$ symbol as a triangle, not a tripod, and the diagram drawn in this way can be shown as a closed hexagon. In contrast, the Yutsis diagram for this setup does not close, implying that our coupling coefficient $\mathcal{G}_{[12]}$, which is a product of six Gaunt integrals, cannot be rewritten as e.g. a 3$n$-$j$ symbol. We also note that in a Yutsis diagram, the thrree inner diagonals here would have to have the same value along their whole lengths; not the case above.}
    \label{fig:angmom0}
\end{figure}

\begin{figure*}
    %\centering
    \includegraphics[width=18cm]{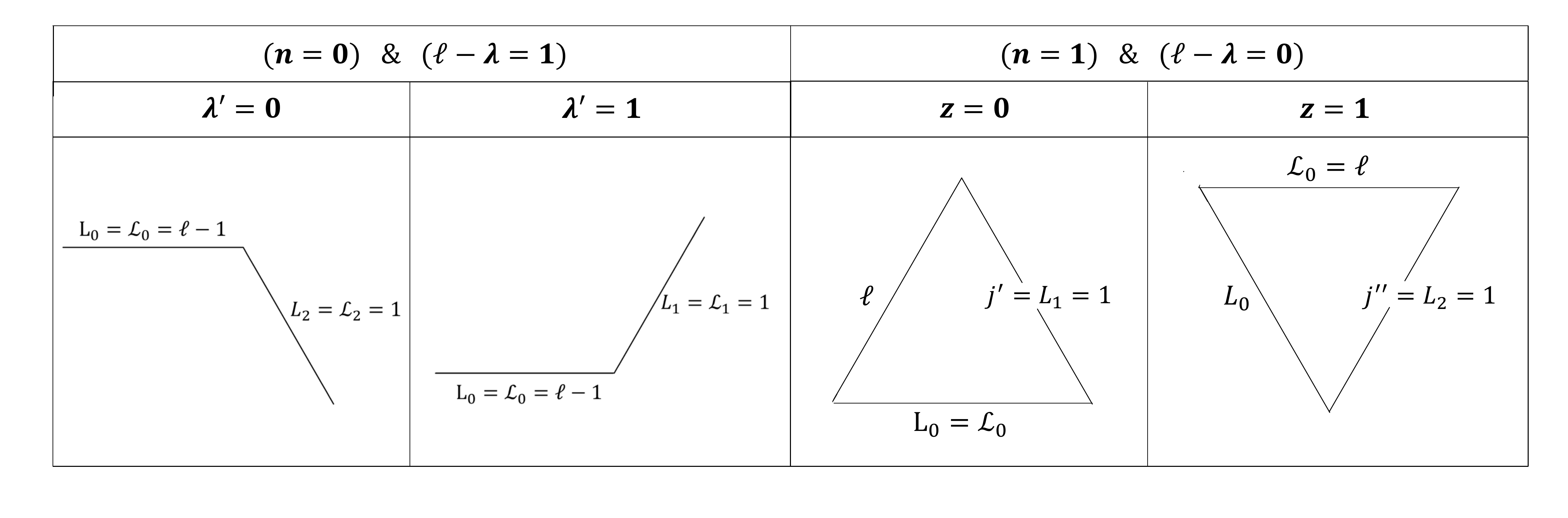}
    \caption{Reduction of Figure \ref{fig:angmom0} considering the cases where $n=0$ and $\ell-\lambda=1$ in the leftmost two panels, and $n=1$ and $\ell-\lambda=0$ in the rightmost two. For the leftmost panels, we consider $\lambda'=0$ and $\lambda'=1$, For the rightmost panels, we consider $z=0$ and $z=1$. These cases are of particular interest because they give the leading-order (i.e. $\epsilon^1$) correction our approach includes beyond the single-triplet-member line of sight. We note that all angular momenta from Figure \ref{fig:angmom0} that are not explicitly shown in the diagrams above are zero. The results of using these diagrams to deduce the spherical harmonic combinations needed for our method's implied leading-order correction to the single-triplet-member line of sight are in equations (\ref{eqn:combo_1}) through (\ref{eqn:combo_4}).}
    \label{fig:n0n1}
\end{figure*}

We summarize the angular momentum structure of $\mathcal{G}_{[12]}$ in Figure \ref{fig:angmom0}, and note that this diagram is general and does not yet assume any reduction of equation (\ref{eqn:main_in_body}). We may then use this diagram in combination with the constraints of equations (\ref{eqn:case_n1}) and (\ref{eqn:case_n0}) to obtain Figure \ref{fig:n0n1}, which summarizes the angular momentum constraints on $L_0, L_1$, and $L_2$ imposed by working at leading order in $\epsilon$.

%%%%

With the aid of Figure \ref{fig:n0n1}, we see that for our first case, where $n=0$, $\ell-\lambda=1$, and in the sub-case $\lambda'=0$, the final spherical harmonics will be
\begin{align}
n= 0,\;\ell - \lambda = 1, \;\lambda' = 0 \rightarrow    Y_{\ell-1}^{M_0}(\hat{x}_1) \; Y_{0}^{0}(\hat{r}_{12}) \;  Y_{1}^{M_2}(\hat{r}_{13}).
\label{eqn:combo_1}
\end{align}
In the sub-case $\lambda'=1$, we have
\begin{align}
n= 0,\;\ell - \lambda = 1, \;\lambda' = 1 \rightarrow     Y_{\ell-1}^{M_0}(\hat{x}_1) \; Y_{1}^{M_1}(\hat{r}_{12}) \;  Y_{0}^{0}(\hat{r}_{13}).
\label{eqn:combo_2}
\end{align}
We now consider the second case, where $n=1$ and $\ell-\lambda=0$, and the sub-case with $z=0$. We have
\begin{align}
n= 1,\;\ell - \lambda = 0, \; z= 0 \rightarrow    Y_{\ell\pm 1}^{m \pm 1}(\hat{x}_1) \; Y_{1}^{M_1}(\hat{r}_{12}) \;  Y_{0}^{0}(\hat{r}_{13})
\label{eqn:combo_3}
\end{align}
For the sub-case where $z = 1$, we have
\begin{align}
n= 1,\;\ell - \lambda = 0, \; z= 1 \rightarrow     Y_{\ell \pm 1}^{M_0}(\hat{x}_1) \; Y_{0}^{0}(\hat{r}_{12}) \;  Y_{1}^{M_2}(\hat{r}_{13}).
\label{eqn:combo_4}
\end{align}
We note that for the $n=1$ cases, the requirement that for a 3-$j$ symbol with zero spins, the upper row's sum must be even, we must have $L_0 = \ell \pm 1$, and $L_0 = \ell$ is not allowed, even though the triangle showing how $L_0$ connects to $\ell$ in Figure \ref{fig:n0n1} would appear to permit this (and indeed, it does satisfy the triangular inequality). 

For $n=0$, $\ell-\lambda=1$, in both the sub-case where $\lambda'=0$ and the  sub-case where $\lambda'=1$ we found that
\begin{align}
    \mathcal{G}_{[12]} = (4\pi)^{-3}.
    \end{align}
For $n=1$, $\ell-\lambda=0$ and $z=0$ we found that
\begin{align}
    \mathcal{G}_{[12]} =& \sqrt{3}(4\pi)^{-3}  \sqrt{(2\ell+1)(2L_0+1)}(-1)^{M_0+M_1}\nonumber\\
&\times    \begin{pmatrix}
    \ell & 1 & L_0\\ 
    \mu & -m' & -\mathcal{M}_0
    \end{pmatrix}
     \begin{pmatrix}
    \ell & 1 & L_0\\ 
    0 & 0 & 0
    \end{pmatrix} .\nonumber
\end{align}
For $n=1$, $\ell-\lambda=0$ and $z=1$:
\begin{align}
    \mathcal{G}_{[12]} =& \sqrt{3} (4\pi)^{-3} (-1)^{\mu+M_2} \sqrt{(2\ell+1)(2L_0+1)}\nonumber\\
   &\times \begin{pmatrix}
    \ell & 1 & L_0\\ 
    \mathcal{M}_0 & -m'' & -M_0
    \end{pmatrix}
     \begin{pmatrix}
    \ell & 1 & L_0\\ 
   0 & 0 & 0
    \end{pmatrix}
    .\nonumber
\end{align}
We now supply the values of $\mathcal{S}$, $\mathcal{U}$, $\mathcal{T}$, and $\mathcal{Q}$ relevant for these leading-order-in-$\epsilon$ cases. 

For $n=0$, $\lambda = \ell - 1$, and both allowed values of $\lambda'$, one needs
\begin{align}
\mathcal{U}^{0 0\; (\ell/2)}_{0 0 0 0 0 0} = 4\pi^2\; \mathcal{T}_{0 0 0}^{0 0 0 \;(\ell/2)} \mathcal{S}_0^0 \mathcal{S}_0^0 = (4\pi)^3.
\end{align}
where we used that
\begin{align}
\mathcal{S}_0^0 = 1,\;\;\; \mathcal{T}_{000}^{000\;(\ell/2)} = 4\pi
\label{eqn:S00}
\end{align}
to obtain the second equality for $\mathcal{U}$ above. 

For $n=1$, $\lambda = \ell$, and $z= 0$, one needs
\begin{align}
\mathcal{U}^{0 1\; (\ell/2)}_{0 0 0 0 1 0} = \frac{4\pi^2}{3}\; \mathcal{T}_{0 0 0}^{0 0 1 \;(\ell/2)} \mathcal{S}_1^1 \mathcal{S}_0^0 = -48\pi^3\ell.
\end{align}
where we used that
\begin{align}
\mathcal{S}_1^1 = \frac{3}{2},\;\;\; \mathcal{T}_{000}^{001\;(\ell/2)} = -24\pi \ell
\label{eqn:S11}
\end{align}
and had computed $\mathcal{S}_0^0$ in equation (\ref{eqn:S00}) above.

For $n=1$, $\lambda = \ell$, and $z= 1$, one needs
\begin{align}
\mathcal{U}^{0 1\; (\ell/2)}_{0 0 0 0 1 0} = \frac{4\pi^2}{3}\; \mathcal{T}_{1 0 0}^{0 0 1 \;(\ell/2)} \mathcal{S}_0^0 \mathcal{S}_1^1 = -48\pi^3\ell
\end{align}
where we used that
\begin{align}
\mathcal{T}_{100}^{001\;(\ell/2)} = -24\pi \ell
\end{align}
and $\mathcal{S}_1^1$ was computed in equation (\ref{eqn:S11}) above.

Finally, for $n=0$, $\lambda' = 0$ and $1$, we have, denoting the two possible values of $\lambda'$ as $[0,1]$ below:
\begin{align}
    &\mathcal{Q}_{\ell (\ell -1) [0,1] } \equiv     
    \begin{pmatrix}
    2\ell\\ 
    2(\ell - 1)
    \end{pmatrix}
    ^{1/2} (-1)^{\ell -1  + \mu} \sqrt{2\ell + 1} \sqrt{3} \nonumber\\
& \times  \begin{pmatrix}
\ell - 1& 1 & \ell\\
\mu & m-\mu & -m
\end{pmatrix} 
\begin{pmatrix}
[0, 1]& 1- [0, 1] & 1\\
\mu' & m - \mu - \mu' & -m + \mu
\end{pmatrix}.
    \nonumber
\end{align} 
We note that the last 3-$j$ symbol for $\lambda' = 1$ is a cyclic permutation of that for $\lambda' = 0$, so in fact $\mathcal{Q}$ has the same overall value for both $\lambda'$ cases.

For $n=1$, $\lambda'= \ell$, $\lambda' =0$, and both values of $z$, we find that
\begin{align}
\mathcal{Q}_{\ell \ell 0} = 1.
\end{align}
With these values in hand, one has the needed weights to apply in summing up the spherical harmonic coefficients of the density field (as it projects onto the spherical harmonics given by equations \ref{eqn:combo_1} through \ref{eqn:combo_4}) into equation (\ref{eqn:main_in_body}).
Thus one has a complete prescription for evaluating the leading-order correction to the anisotropic 3PCF estimate to go beyond the single-triplet-member line of sight and use either the centroid or unit-vector-average methods proposed in this work.

\iffalse
\section{Edge correction}
\label{sec:edgecorr}

{\bf ZS: I think this proceeds in the same way as in our earlier paper. I may write this.}
\fi

\section{Discussion and Conclusions}
\label{sec:conclusions}
In this work, we have for the first time shown how to evaluate a fully symmetric line of sight to a triplet of galaxies yet still have an algorithm that scales as $N^2$, with $N$ the number of objects. We have obtained an exact expansion of the line of sight defined either as the average of the three position vectors to a triplet, or the average of the three direction vectors, and also reduced this complicated expression to a compact set of leading-order corrections one would compute beyond what is already present in the single-triplet-member line-of-sight estimator. We have shown that, if one wants the $\ell^{\th}$ harmonic moment of the line of sight $\hat{n}$, one can obtain the leading-order correction relative to the single-triplet-member approach using just three additional total angular momentum combinations (see equations \ref{eqn:combo_1} through \ref{eqn:combo_4}). We need only three because, due to the interchange symmetry of this set, there are only three distinct combinations; the $\hat{r}_{ij}$ are dummy variables to be integrated over to get the density coefficients. Given that, according to \cite{Sugiyama_2018}, most of the signal is expected in $\ell = 2$ and $4$, leading to $\ell + 1 = 5$ at most, not many spins would be needed (just 6 for the $\ell = 5$ harmonic implied by the larger $\ell$ in equation \ref{eqn:combo_3}). Similarly, there are only two spins possible for the harmonics of $\hat{r}_{ij}$ there. Furthermore, due to the factorization of the angular integrals, they can be considered separately, so the computational cost does not multiply.

Given the significant improvement on cosmological parameters forecast for anisotropic 3PCF by a number of the works mentioned in \S\ref{sec:intro} , as well as the rich upcoming data landscape of spectroscopic surveys such as DESI, Roman, and Euclid, tools for the precise measurement of triplet clustering in redshift space are vital. \cite{Sugiyama_2018} showed that the line of sight definition can make a difference that, while negligible for BOSS, would be quite significant for DESI. Here, we have developed the tool to address this challenge.

Much future work remains to be done. First will be implementing the algorithm of this work. In this context, harnessing some of the accelerations for spherical harmonic computation developed in \cite{SE_3pt_alg} and \cite{Friesen_2017} will be particularly valuable. We also note that, much as shown in \cite{SE_3pt_FT}, the harmonic coefficients can be obtained using Fast Fourier Transforms, enabling an even faster scaling ($N_{\rm g} \log N_{\rm g}$) than $N^2$. Another issue to be addressed in future work is how edge correction should be done in our proposed basis. Presumably an angular-momentum coupling matrix analogous to those used in \cite{SE_3pt_alg} and \cite{SE_aniso_3PCF} can be derived, but this remains to be done.

Overall, however, we hope already in its present form this work is a useful first step towards bringing to the anisotropic 3PCF the same precision and accuracy that has long been demanded for anisotropic 2PCF. We also note that, much as the basis of \cite{Sugiyama_2018} is agnostic as to whether one works in configuration space or Fourier space, our method here can be seamlessly employed to estimate Fourier-space anisotropic bispectrum as well. 

Finally, we note that one area where one considers quite large scales, leading to large angles subtended at the observer, and also particularly desires to harness the 3PCF, is primordial non-Gaussianity (PNG). Here, the signal expected is quite small and hence total control of systematics, including any errors from line-of-sight definition, will be vital. With possible future missions that will focus on PNG (e.g. \citealt{schlegel2019}), new tools to maximize the value extracted from them are worth pursuing fully.

%%%%%%%%%%%%%%%%%%%%%%%%%%%%%%%%%%%%%%%%%%%%%%%%%%

%%%%%%%%%%%%%%% ACKNOWLEDGEMENTS %%%%%%%%%%%%%%%%%
\section*{Acknowledgements}
KG thanks Farshad Kamalinejad and Naman Shukla for useful discussions. ZS thanks Alex Krolewski, Kristen Lavelle, and Stephen Portillo for useful discussions.

%%%%%%%%%%%%%%%%%%%%%%%%%%%%%%%%%%%%%%%%%%%%%%%%%%

%%%%%%%%%%%%%%%%% APPENDICES %%%%%%%%%%%%%%%%%%%%%
\appendix
\section{Use of Solid Harmonic Shift Theorem}
\label{app:app1}
Here we derive the expansion of $Y_{\ell}^{m}(\reallywidehat{3\vec{x}_1 + \vec{r}_{12} + \vec{r}_{13}})$ in terms of a sum of triple products of single-argument spherical harmonics $Y_{\ell}^{m}$ of $\hat{x}_1$, $\hat{r}_{12}$ and $\hat{r}_{13}$. We use $\vec{d} \equiv \vec{r}_{12} + \vec{r}_{13}$ as an auxiliary variable to simplify some steps of our calculation. To derive the expansion we will exploit the solid harmonics
\begin{align}
    \label{eqn:harmonics}
    R_\ell^m(\vec{r}) \equiv \sqrt{\frac{4\pi}{2\ell+1}} r^\ell Y_{\ell}^{m}(\hat{r}),
\end{align}
and the solid harmonic shift theorem
\begin{align}
    R_\ell^m(\vec{r}+\vec{s}) &= \sum_{\lambda=0}^{\ell}
    \begin{pmatrix}
    2\ell\\ 
    2\lambda
    \end{pmatrix}
    ^{1/2}
    \sum_{\mu=-\lambda}^{\lambda} R_\lambda^\mu (\vec{r}) R_{\ell-\lambda}^{m-\mu} (\vec{s})\nonumber\\
    &\times \left<\lambda,\mu; \ell-\lambda,m-\mu|\ell m \right>,
    \label{eq:shift_thm}
\end{align}
which we prove by explicit calculation for the cases $\ell = \{0,1,2\}$ in Appendix \ref{app:app2}. On the righthand side above, the term in parentheses immediately following the sum over $\lambda$ is a binomial coefficient and the last line is a Clebsch-Gordan coefficient, which can be related to 3-$j$ symbols. We will do this in our calculation shortly, but state the theorem as above to accord with the form in which it standardly appears.

We will first compute $Y_{\ell}^{m}(\reallywidehat{3\vec{x}_1+\vec{d}})$ using the shift theorem. Then we will substitute $\vec{r}_{12}$ and $\vec{r}_{13}$ back in for $\vec{d}$ and use the shift theorem again. Applying equation (\ref{eq:shift_thm}) to $R_\ell^m(3\vec{x}_1+\vec{d})$ we find
\begin{align}
    \label{eqn:shift_thm}
    R_\ell^m(3\vec{x}_1+\vec{d}) &= \sum_{\lambda=0}^{\ell}
    \begin{pmatrix}
    2\ell\\ 
    2\lambda
    \end{pmatrix}
    ^{1/2} \sum_{\mu=-\lambda}^{\lambda} R_\lambda^\mu (3\vec{x}_1) R_{\ell-\lambda}^{m-\mu} (\vec{d})\nonumber\\
    &\times \left<\lambda,\mu; \ell-\lambda,m-\mu|\ell m \right>.
\end{align}
Substituting equation (\ref{eqn:harmonics}) for the solid harmonics on the right-hand side above and rearranging, we obtain:
\begin{align}
    \label{eqn:intermediate_result}
     Y_{\ell}^m &(\reallywidehat{3\vec{x}_1+\vec{d}}) = \\
    &\sqrt{\frac{2\ell+1}{4\pi}} \left|\;3\vec{x}_1 + \vec{d\;} \right|^{-\ell} \sum_{\lambda=0}^{\ell} \begin{pmatrix}
    2\ell\\ 
    2\lambda
    \end{pmatrix}
    ^{1/2} \sum_{\mu=-\lambda}^{\lambda} \sqrt{\frac{4\pi}{2\lambda + 1}} (3x_1)^\lambda Y_\lambda^\mu(\hat{x}_1) \nonumber\\
    &\times \sqrt{\frac{4\pi}{2(\ell - \lambda) + 1}} d^{\ell - \lambda} Y_{(\ell - \lambda)}^{(m - \mu)}(\hat{d}\;) \left<\lambda,\mu; \ell - \lambda, m - \mu | \ell m \right>.\nonumber
\end{align}
We now need to expand the spherical harmonic of $\hat{d}$ on the righthand side above, since $\vec{d} = \vec{r}_{12} + \vec{r}_{13}$; we use the shift theorem again to do so. Simplifying what results, we obtain
\begin{align}
    \label{eqn:fac_interm}
    Y_\ell^m &(\reallywidehat{3\vec{x}_1 + \vec{r}_{12} + \vec{r}_{13}}) =  \frac{4\pi}{|3\vec{x}_1+\vec{r}_{12}+\vec{r}_{13}|^\ell} \sum_{\lambda=0}^{\ell} \sum_{\lambda'=0}^{\ell-\lambda} \mathcal{Q}_{\ell \lambda \lambda'} \nonumber\\
    &\times (3x_1)^\lambda r_{12}^{\lambda'} r_{13}^{\ell-\lambda-\lambda'} \nonumber \\
    &\times \sum_{\mu = - \lambda}^{\lambda} \sum_{\mu' = -\lambda'}^{\lambda'}\; Y_\lambda^\mu(\hat{x}_1)\; Y_{\lambda'}^{\mu'}(\hat{r}_{12})\; Y_{(\ell-\lambda-\lambda')}^{(m-\mu -\mu')}(\hat{r}_{13}) ,
\end{align}
where
\iffalse
\begin{align}
    \mathcal{Q}_{\ell \lambda \lambda'} &= \begin{pmatrix}
    2\ell\\ 
    2\lambda
    \end{pmatrix}
    ^{1/2}
    \begin{pmatrix}
    2(\ell-\lambda)\\ 
    2\lambda'
    \end{pmatrix}
    ^{1/2} \\ 
    &\times \sqrt{\frac{(2\ell+1)[2(\ell-\lambda)+1]}{(2\lambda+1)[2(\ell-\lambda)+1](2\lambda'+1)[2(\ell-\lambda-\lambda')+1]}}.
    \nonumber
\end{align}
\fi
\begin{align}
    \mathcal{Q}_{\ell \lambda \lambda'} &= \begin{pmatrix}
    2\ell\\ 
    2\lambda
    \end{pmatrix}
    ^{1/2}
    \begin{pmatrix}
    2(\ell-\lambda)\\ 
    2\lambda'
    \end{pmatrix}^{1/2} \left< \lambda, \mu; \ell - \lambda, m - \mu | \ell m \right>\\ 
    &\times  \left< \lambda', \mu'; \ell - \lambda - \lambda', m - \mu - \mu' | \ell - \lambda, m - \mu \right> \nonumber\\
    & =     \begin{pmatrix}
    2\ell\\ 
    2\lambda
    \end{pmatrix}
    ^{1/2}
    \begin{pmatrix}
    2(\ell-\lambda)\\ 
    2\lambda'
    \end{pmatrix}
    ^{1/2} (-1)^{\lambda + \mu} \sqrt{2\ell + 1} \sqrt{2(\ell - \lambda) + 1} \nonumber\\
& \times  \begin{pmatrix}
\lambda & \ell - \lambda & \ell\\
\mu & m-\mu & -m
\end{pmatrix} \begin{pmatrix}
\lambda' & \ell - \lambda - \lambda' & \ell - \lambda\\
\mu' & m - \mu - \mu' & -m + \mu
\end{pmatrix}.
    \nonumber
\end{align}
We used the relation between Clebsch-Gordan symbols and 3-$j$ symbols to convert to the latter in the second equality.\footnote{E.g. \url{https://mathworld.wolfram.com/Clebsch-GordanCoefficient.html} equation (7).}
Equation (\ref{eqn:fac_interm}) is nearly fully factorized in $3\vec{x}_1$, $\vec{r}_{12}$, and $\vec{r}_{13}$; however we have the pre-factor $|3\vec{x}_1 + \vec{r}_{12} + \vec{r}_{13}|^{-\ell}$. We now consider how to expand this term.

For $\ell = 1$, and reverting to our auxiliary variable $\vec{d} = \vec{r}_{12} + \vec{r}_{13}$, this pre-factor would be
\begin{align}
\frac{1}{|3\vec{x}_1 + \vec{d}|} = \frac{1}{3x_1}\sum_{n = 0}^{\infty} \left(\frac{d}{3x_1}\right)^n \mathcal{L}_{n}(\hat{x}_1 \cdot\hat{d}),
\end{align}
with $\mathcal{L}_n$ the Legendre polynomial of order $n$. However, we need more general powers beyond just $\ell = 1$. Pulling out $(3x_1)^{\ell}$ we find
\begin{align}
\frac{1}{|3\vec{x}_1 + \vec{d}|^\ell} = \frac{1}{(3x_1)^{\ell}}\frac{1}{|\hat{x}_1 + [d/(3x_1)] \hat{d}|^{\ell}}.
\end{align}
Rewriting the magnitude in the denominator on the righthand side above as a square-root, we find
\begin{align}
\label{eqn:fac_interm_l}
\frac{1}{|3\vec{x}_1 + \vec{d}|^\ell} &=\nonumber\\ &\frac{1}{(3x_1)^{\ell}}\frac{1}{\sqrt{1 + 2[d/(3x_1)] (\hat{x}_1 \cdot \hat{d}) +  [d/(3x_1)]^2}^{\ell}}.
\end{align}
Now we see that equation (\ref{eqn:fac_interm_l}) has the form
\begin{align}
   \label{eqn:gegenbauer_12}
   \frac{1}{|3\vec{x}_1 + \vec{d}|^\ell} = \frac{1}{(3x_1)^{\ell}} \frac{1}{(1 - 2 y t + t^2 )^{\alpha}}
\end{align}
with $ t = -d/(3x_1) $, $\alpha = \ell /2$, and $y = \hat{x}_1 \cdot \hat{d}$.
Equation (\ref{eqn:gegenbauer_12}) in turn is a rescaling (by $(3x_1)^{-\ell}$) of the generating function for Gegenbauer polynomials $C_n^{(\alpha)}(y)$, so we may expand it as
\begin{align}
    \label{eqn:gegenbauer_13}
   \frac{1}{|3\vec{x}_1 + \vec{d}|^\ell} = \frac{1}{(3x_1)^{\ell}} \sum_{n=0}^\infty C_n^{(\ell/2)} (y)\;t^n.
\end{align}
To proceed further, we will focus on manipulating  $\Cn(y)\;t^n$ and use the explicit form of the Gegenbauer polynomials,
\begin{align}
C_n^{(\alpha)}(y)   = \sum_{k = 0}^{\lceil n/2\rceil} C_{n k}^{\alpha} y^{n-2k}
\end{align}
where
\begin{align}
   C_{n k}^{(\alpha)} \equiv (-1)^k (2)^{n-2k} \frac{\Gamma(n-k+\alpha)}{\Gamma(\alpha)k!(n-2k)!}
\end{align}
is an expansion coefficient. $\lceil n/2\rceil$ means the largest integer less than or equal to $n/2$. Making the appropriate replacements for $y$ and for $t$, we now find that
\begin{align}
    t^n &C_n^{(\alpha)} (y) \\
    &= \sum_{k=0}^{\lfloor n/2 \rfloor} C_{n k}^{(\alpha)} \left( 2 \hat{x}_1 \cdot \frac{\vec{r}_{12} + \vec{r}_{13}}{|\vec{r}_{12} + \vec{r}_{13}|} \right)^{n-2k}(-1)^n \left( \frac{|\vec{r}_{12} + \vec{r}_{13}|}{3x_1} \right)^n. \nonumber \\
    &= \sum_{k=0}^{\lfloor n/2 \rfloor} C_{n k}^{(\alpha)} \left( 2 \hat{x}_1 \cdot [\vec{r}_{12} + \vec{r}_{13}] \right)^{n-2k} |\vec{r}_{12} + \vec{r}_{13}|^{2k} \left(- \frac{1}{3x_1} \right)^n.
    \nonumber
\end{align}
As defined before, $\epsilon_{12,1} \equiv r_{12}/(3 x_1)$ and $\epsilon_{13,1} \equiv r_{13}/(3 x_1)$, so we can rewrite the above equation as
\begin{align}
    \label{tnCn}
    t^n C_n^{(\alpha)} (y) &= \sum_{k=0}^{\lfloor n/2 \rfloor} C_{n k}^\alpha \left( -\frac{1}{x_1} \right)^n \\
    & \times (2x_1)^{n-2k} [\epsilon_{12,1} \; (\hat{x}_1 \cdot \hat{r}_{12}) + \epsilon_{13,1} \; (\hat{x}_1 \cdot \hat{r}_{13})]^{n-2k} \nonumber \\
    & \times x_1^{2k} [\epsilon_{12,1}^2 + \epsilon_{13,1}^2 + 2\epsilon_{12,1}\epsilon_{13,1} (\hat{r}_{12} \cdot \hat{r}_{13})]^k
    \nonumber
\end{align}
In equation (\ref{tnCn}), there are two sums raised to powers, each of which must be expanded. The first one becomes
\begin{align}
    [\epsilon_{12,1} \; (\hat{x}_1 \cdot &\hat{r}_{12}) \; + \; \epsilon_{13,1} \; (\hat{x}_1 \cdot \hat{r}_{13})]^{n-2k} = \\ &\sum_{z=0}^{n-2k}
    \begin{pmatrix}
    n-2k\\
    z
    \end{pmatrix}
    (\epsilon_{12,1} \; \mu_{12,1})^{n-2k-z} (\epsilon_{13,1} \; \mu_{13,1})^{z} ,
    \nonumber
\end{align}
where $\mu_{ij,k} \equiv \hat{r}_{ij} \cdot \hat{x}_k$. By two successive applications of the binomial theorem, the second term (last line of equation \ref{tnCn}) becomes
\begin{align}
    [&\epsilon_{12,1}^2 + \epsilon_{131}^2 + 2\;\epsilon_{12,1}\;\epsilon_{13,1} \; (\hat{r}_{12} \cdot \hat{r}_{13})]^k \\
    &= \sum_{z'=0}^{k}
    \begin{pmatrix}
    k\\
    z'
    \end{pmatrix}
    (\epsilon_{12,1}^2 + \epsilon_{13,1}^2)^{k-z'} (2 \; \epsilon_{12,1}\;\epsilon_{13,1} \hat{r}_{12} \cdot \hat{r}_{13})^{z'} \nonumber \\
    &= \sum_{z'=0}^{k}
    \begin{pmatrix}
    k\\
    z'
    \end{pmatrix}
    \sum_{z''=0}^{k-z'}
    \begin{pmatrix}
    k-z'\\
    z''
    \end{pmatrix}  \epsilon_{12,1}^{2(k-z'-z'')} \epsilon_{13,1}^{2z''} (2\epsilon_{12,1}\epsilon_{13,1} \hat{r}_{12} \cdot \hat{r}_{13})^{z'} \nonumber \\
    &= \sum_{z'=0}^{k}
    \begin{pmatrix}
    k\\
    z'
    \end{pmatrix}
    \sum_{z''=0}^{k-z'}
    \begin{pmatrix}
    k-z'\\
    z''
    \end{pmatrix}
    \epsilon_{12,1}^{2(k-z'-z'')} \epsilon_{13,1}^{2z''} 2^{z'} \epsilon_{12,1}^{z'} \epsilon_{13,1}^{z'} (\hat{r}_{12} \cdot \hat{r}_{13})^{z'} .
    \nonumber
\end{align}

Now we have yet another term which can be expanded into Legendre polynomials:
\begin{align}
    \label{eqn:lll}
    (\hat{r}_{12} \cdot &\hat{r}_{13})^{z'} \\
    &= \sum_{j=z'\downarrow 2} \mathcal{S}_j^{z'} \; P_j(\hat{r}_{12} \cdot \hat{r}_{13}) \nonumber \\
    &= \sum_{j=z'\downarrow 2} \mathcal{S}_j^{z'} \; \frac{4\pi}{2j+1} \sum_{m=-j}^{j}Y_j^m(\hat{r}_{12})Y*_{jm}(\hat{r}_{13}) ,
    \nonumber
\end{align}
where we have defined
\begin{align}
    \mathcal{S}_j^{z'} \equiv \frac{(2j+1)z'!}{2^{(z'-j)/2}((z'-j)/2)! (j+z'+1)!!}.
\end{align}
Putting all these together, we obtain the final form of the Gegenbauer polynomial we are interested in:
\begin{align}
    t^n C_n^{(\alpha)} (y)& = \sum_{k=0}^{\lfloor n/2 \rfloor} C_{n k}^\alpha (2x_1)^{n-2k} \left( -\frac{1}{x_1} \right)^n \\
    & \times \sum_{z=0}^{n-2k}
    \begin{pmatrix}
    n-2k\\
    z
    \end{pmatrix}
    (\epsilon_{12,1} \mu_{12,1})^{n-2k-z} (\epsilon_{13,1} \mu_{13,1})^{z} \; x_1^{2k} \nonumber \\
    & \times \sum_{z'=0}^{k}
    \begin{pmatrix}
    k\\
    z'
    \end{pmatrix}
    \sum_{z''=0}^{k-z'}
    \begin{pmatrix}
    k-z'\\
    z''
    \end{pmatrix}
    \epsilon_{12,1}^{2(k-z'-z'')} \epsilon_{13,1}^{2z''} 2^{z'} \epsilon_{12,1}^{z'} \epsilon_{13,1}^{z'} \nonumber \\
    & \times \sum_{j=z'\downarrow 2} \mathcal{S}_j^{z'} \; \frac{4\pi}{2j+1} \sum_{m=-j}^{j}Y_j^m(\hat{r}_{12})Y^*_{jm}(\hat{r}_{13}) .
    \nonumber
\end{align}
Rearranging and cancelling out terms we obtain
\begin{align}
    \label{eqn:tnCn_4}
    t^n C_n^{(\alpha)} (y) = &\sum_{k=0}^{\lfloor n/2 \rfloor} \; \sum_{z=0}^{n-2k} \; \sum_{z'=0}^{k} \; \sum_{z''=0}^{k-z'} \; \sum_{j=z'\downarrow 2} \;  \sum_{m=-j}^{j} \nonumber \\
    & \mathcal{T}_{z z'z''}^{j k n \alpha} \; Y_j^m(\hat{r}_{12})\; Y_j^{m*}(\hat{r}_{13}) \\
    & \times \epsilon_{12,1}^{n-z-z'-2z''} \; \epsilon_{13,1}^{z+z'+2z''} \; \mu_{12,1}^{n-2k-z} \; \mu_{13,1}^z ,
    \nonumber
\end{align}
where we have defined
\begin{align}
    &\mathcal{T}_{z z' z''}^{j k n \alpha} \equiv \frac{\Gamma(n-k+\alpha)}{\Gamma(\alpha)}\\
    &\times \frac{4\pi \; (-1)^k \; (2)^{z'+2n-4k-(z'-j)/2}}{z! \; z''! \; (n-2k-z)! \; (k-z'-z'')! \; [(z'-j)/2]! \; (j+z'+1)!!}
    \nonumber
\end{align}
We now expand the $\mu$ terms in equation (\ref{eqn:tnCn_4}) into Legendre polynomials. First,
\begin{align}
    \mu_{12,1}^{n-2k-z} &= \sum_{j'=(n-2k-z) \downarrow 2} \mathcal{S}_{j'}^{n-2k-z} \; P_{j'} (\hat{r}_{12}\cdot\hat{x_1}) \nonumber \\
    &= \sum_{j'=(n-2k-z) \downarrow 2} \mathcal{S}_{j'}^{n-2k-z} \nonumber \\ &\times \frac{4\pi}{2j'+1} \sum_{m'=-j'}^{j'} Y_{j'}^{m'}(\hat{r}_{12}) Y_{j'}^{m'*} (\hat{x_1}) .
\end{align}
Then
\begin{align}
    \mu_{13,1}^{z} &= \sum_{j''=z \downarrow 2} \mathcal{S}_{j''}^{z} \; P_{j''} (\hat{r}_{13}\cdot\hat{x_1}) \nonumber \\
    &= \sum_{j''=z \downarrow 2} \mathcal{S}_{j''}^{z} \nonumber \\
    &\times \frac{4\pi}{2j''+1} \sum_{m''=-j''}^{j''} Y_{j''}^{m''}(\hat{r}_{13}) Y_{j''}^{m''*} (\hat{x_1}) .
\end{align}
Substituting the above expansions into equation (\ref{eqn:tnCn_4}), we obtain
\begin{align}
    t^n &C_n^{(\alpha)} (y) = \sum_{k=0}^{\lfloor n/2 \rfloor} \; \sum_{z=0}^{n-2k} \; \sum_{z'=0}^{k} \; \sum_{z''=0}^{k-z'} \; \sum_{j=z'\downarrow 2} \;  \sum_{m=-j}^{j} \nonumber \\
    & \sum_{j'=(n-2k-z) \downarrow 2} \; \sum_{j''=z \downarrow 2} \; \mathcal{T}_{z z' z''}^{j k n \alpha} \nonumber\\ 
    & \times Y_j^m(\hat{r}_{12})\; Y_j^{m*}(\hat{r}_{13}) \epsilon_{12,1}^{n-z-z'-2z''} \; \epsilon_{13,1}^{z+z'+2z''} \\
    & \times \mathcal{S}_{j'}^{n-2k-z} \; \frac{4\pi}{2j'+1} \sum_{m'=-j'}^{j'} Y_{j'}^{m'}(\hat{r}_{12}) Y_{j'}^{m'*} (\hat{x_1}) \nonumber \\
    & \times \mathcal{S}_{j''}^{z} \; \frac{4\pi}{2j''+1} \sum_{m''=-j''}^{j''} Y_{j''}^{m''}(\hat{r}_{13}) Y_{j''}^{m''*} (\hat{x_1}) .
    \nonumber
\end{align}
Simplifying the coefficients we find
\begin{align}
    \label{eqn:tnCn_5}
    t^n &C_n^{(\alpha)} (y) = \sum_{k=0}^{\lfloor n/2 \rfloor} \; \sum_{z=0}^{n-2k} \; \sum_{z'=0}^{k} \; \sum_{z''=0}^{k-z'} \; \sum_{j=z'\downarrow 2} \;  \sum_{m=-j}^{j} \nonumber \\
    & \sum_{j'=(n-2k-z) \downarrow 2} \sum_{m'=-j'}^{j'} \; \sum_{j''=z \downarrow 2} \; \sum_{m''=-j''}^{j''} \nonumber \\
    & \times \mathcal{U}_{z z' z'' j j' j''}^{k n \alpha} \nonumber\\ 
    & \times Y_j^m(\hat{r}_{12})\; Y_j^{m*}(\hat{r}_{13}) \; \epsilon_{12,1}^{n-z-z'-2z''} \; \epsilon_{13,1}^{z+z'+2z''} \\
    & \times Y_{j'}^{m'}(\hat{r}_{12}) \; Y_{j'}^{m'*} (\hat{x_1}) \; Y_{j''}^{m''}(\hat{r}_{13}) \; Y_{j''}^{m''*} (\hat{x_1}) ,
    \nonumber
\end{align}
where we have defined
\begin{align}
    \mathcal{U}_{z z' z'' j j' j''}^{k n \alpha} \equiv \frac{4\pi^2}{(2j'+1)(2j''+1)} \; \mathcal{T}_{z z' z''}^{j k n \alpha} \; \mathcal{S}_{j'}^{n-2k-z} \; \mathcal{S}_{j''}^{z} .
\end{align}
Substituting equation (\ref{eqn:gegenbauer_13}) in equation (\ref{eqn:fac_interm}) we find
\begin{align}
    \label{eqn:fac_interm_2}
    Y_\ell^m &(\reallywidehat{3\vec{x}_1 + \vec{r}_{12} + \vec{r}_{13}}) =  \frac{4\pi}{(3x_1)^\ell} \sum_{n=0}^\infty C_n^{(\ell/2)} (y) t^n \nonumber \\
    & \hspace{1cm} \times \sum_{\lambda=0}^{\ell} \; \sum_{\lambda'=0}^{\ell-\lambda} \mathcal{Q}_{\ell \lambda \lambda'}\; (3x_1)^\lambda \; r_{12}^{\lambda' }\; r_{13}^{\ell-\lambda-\lambda'} \\
    & \hspace{1cm} \times \sum_{\mu = - \lambda}^{\lambda} \; \sum_{\mu' = -\lambda'}^{\lambda'}\; Y_\lambda^\mu(\hat{x}_1)\; Y_{\lambda'}^{\mu'}(\hat{r}_{12})\; Y_{(\ell-\lambda-\lambda')}^{(m-\mu -\mu')}(\hat{r}_{13}).\nonumber
\end{align}
We note that we have now set $\alpha = \ell/2$ in the Gegenbauer polynomial as appropriate for our desired case. Now substituting equation (\ref{eqn:tnCn_5}) in equation (\ref{eqn:fac_interm_2}) we obtain
\begin{align}
    \label{eqn:almostthere}
    Y_\ell^m &(\reallywidehat{3\vec{x}_1 + \vec{r}_{12} + \vec{r}_{13}}) =  \frac{4\pi}{(3x_1)^\ell} \sum_{n=0}^\infty \sum_{k=0}^{\lfloor n/2 \rfloor} \; \sum_{z=0}^{n-2k} \; \sum_{z'=0}^{k} \; \sum_{z''=0}^{k-z'} \nonumber \\
    & \times \sum_{j=z'\downarrow 2}\sum_{m=-j}^j \; \sum_{j'=(n-2k-z) \downarrow 2} \; \sum_{m'=-j'}^{j'} \; \sum_{j''=z \downarrow 2} \; \sum_{m''=-j''}^{j''} \nonumber \\
    & \times \sum_{\lambda=0}^{\ell} \; \sum_{\lambda'=0}^{\ell-\lambda} \; \sum_{\mu = - \lambda}^{\lambda} \; \sum_{\mu' = -\lambda'}^{\lambda'} \; \mathcal{U}_{z z' z'' j j' j''}^{k n\; (\ell/2)}  \; Q_{\ell \lambda \lambda'} \\ 
    & \times \epsilon_{12,1}^{n-z-z'-2z''} \; \epsilon_{13,1}^{z+z'+2z''} \; (3x_1)^{\lambda} \; r_{12}^{\lambda'}\; r_{13}^{\ell-\lambda-\lambda'} \nonumber \\
    & \times Y_{\lambda}^{\mu}(\hat{x}_1) \; Y_{j'}^{m'*} (\hat{x_1}) \; Y_{j''}^{m''*} (\hat{x}_1) \; Y_{\lambda'}^{\mu'}(\hat{r}_{12}) \; Y_j^m(\hat{r}_{12}) \; Y_{j'}^{m'}(\hat{r}_{12}) \nonumber \\
    & \times Y_{(\ell-\lambda-\lambda')}^{(m-\mu -\mu')}(\hat{r}_{13}) \; Y_j^{m*}(\hat{r}_{13}) \; Y_{j''}^{m''}(\hat{r}_{13}) .
    \nonumber
\end{align}
Using the results obtained in Appendix \ref{app:app3}, we turn the product of spherical harmonics left in equation (\ref{eqn:almostthere}) into a sum of expansions:
\begin{align}
    \label{eqn:useappc}
    &Y_\lambda^\mu(\hat{x}_1) \; Y_{j'}^{m'*} (\hat{x}_1) \; Y_{j''}^{m''*} (\hat{x}_1) = \nonumber \\
    & \hspace{1cm} \sum_{L_0 M_0} \; \sum_{\mathcal{L}_0 \mathcal{M}_0} \; \mathcal{G}_{\lambda j' \mathcal{L}_0}^{\mu -m' -\mathcal{M}_0} \; \mathcal{G}_{\mathcal{L}_0 j'' L_0}^{\mathcal{M}_0 -m'' -M_0} \; Y_{L_0}^{M_0}(\hat{x}_1)
    \nonumber \\ \nonumber \\
    &Y_{\lambda'}^{\mu'}(\hat{r}_{12}) \; Y_{j}^{m} (\hat{r}_{12}) \; Y_{j'}^{m'} (\hat{r}_{12}) = \\
    & \hspace{1cm} \sum_{L_1 M_1} \; \sum_{\mathcal{L}_1 \mathcal{M}_1} \; \mathcal{G}_{\lambda' j \mathcal{L}_1}^{\mu' m -\mathcal{M}_1} \; \mathcal{G}_{\mathcal{L}_1 j' L_1}^{\mathcal{M}_1 m' -M_1} \; Y_{L_1}^{M_1}(\hat{r}_{12})
    \nonumber \\ \nonumber \\
    &Y_{(\ell-\lambda-\lambda')}^{(m-\mu -\mu')}(\hat{r}_{13}) \; Y_j^{m*}(\hat{r}_{13}) \; Y_{j''}^{m''}(\hat{r}_{13}) = \nonumber \\
    & \hspace{1cm} \sum_{L_2 M_2} \; \sum_{\mathcal{L}_2 \mathcal{M}_2} \; \mathcal{G}_{(\ell-\lambda-\lambda') j \mathcal{L}_2}^{(m-\mu-\mu') -m -\mathcal{M}_2} \; \mathcal{G}_{\mathcal{L}_2 j'' L_2}^{\mathcal{M}_2 m'' -M_2} \; Y_{L_2}^{M_2}(\hat{r}_{13}) .
    \nonumber
\end{align}
Substituting equations (\ref{eqn:useappc}) into equation (\ref{eqn:almostthere}) we find that
\begin{align}
    Y_\ell^m &(\reallywidehat{3\vec{x}_1 + \vec{r}_{12} + \vec{r}_{13}}) =  \frac{4\pi}{(3x_1)^\ell} \sum_{n=0}^\infty \sum_{k=0}^{\lfloor n/2 \rfloor} \; \sum_{z=0}^{n-2k} \; \sum_{z'=0}^{k} \; \sum_{z''=0}^{k-z'} \nonumber \\
    & \times \sum_{j=z'\downarrow 2} \; \sum_{m=-j}^{j} \; \sum_{j'=(n-2k-z) \downarrow 2} \; \sum_{m'=-j'}^{j'} \; \sum_{j''=z \downarrow 2} \; \sum_{m''=-j''}^{j''} \; \sum_{\lambda=0}^{\ell} \nonumber \\
    & \times \sum_{\lambda'=0}^{\ell-\lambda} \; \sum_{\mu = - \lambda}^{\lambda} \; \sum_{\mu' = -\lambda'}^{\lambda'} \; \sum_{L_0 M_0} \; \sum_{\mathcal{L}_0 \mathcal{M}_0} \; \sum_{L_1 M_1} \; \sum_{\mathcal{L}_1 \mathcal{M}_1} \; \sum_{L_2 M_2} \; \sum_{\mathcal{L}_2 \mathcal{M}_2} \nonumber\\
    & \times \mathcal{U}_{z z' z'' j j' j''}^{k n \;(\ell/2)}  \; Q_{\ell \lambda \lambda'} \nonumber \\
    & \times \mathcal{G}_{\lambda j' \mathcal{L}_0}^{\mu -m' -\mathcal{M}_0} \; \mathcal{G}_{\mathcal{L}_0 j'' L_0}^{\mathcal{M}_0 -m'' -M_0} \; \mathcal{G}_{\lambda' j \mathcal{L}_1}^{\mu' m -\mathcal{M}_1} \nonumber \\
    & \times \mathcal{G}_{\mathcal{L}_1 j' L_1}^{\mathcal{M}_1 m' -M_1} \; \mathcal{G}_{(\ell-\lambda-\lambda') j \mathcal{L}_2}^{(m-\mu-\mu') -m -\mathcal{M}_2} \; \mathcal{G}_{\mathcal{L}_2 j'' L_2}^{\mathcal{M}_2 m'' -M_2} \\ 
    & \times \epsilon_{12,1}^{n-z-z'-2z''} \; \epsilon_{13,1}^{z+z'+2z''} \; (3x_1)^{\lambda} \; r_{12}^{\lambda'} \; r_{13}^{\ell-\lambda-\lambda'} \nonumber \\
    & \times Y_{L_0}^{M_0}(\hat{x}_1) Y_{L_1}^{M_1}(\hat{r}_{12}) Y_{L_2}^{M_2}(\hat{r}_{13}) .
    \nonumber
\end{align}
Substituting $r_{12} = 3\epsilon_{12,1} x_1$, and $r_{13} = 3\epsilon_{13,1} x_1$ we finally obtain
\begin{align}
    Y_\ell^m &(\reallywidehat{3\vec{x}_1 + \vec{r}_{12} + \vec{r}_{13}}) =  4\pi \sum_{n=0}^\infty \sum_{k=0}^{\lfloor n/2 \rfloor} \; \sum_{z=0}^{n-2k} \; \sum_{z'=0}^{k} \; \sum_{z''=0}^{k-z'} \; \sum_{j=z'\downarrow 2} \nonumber \\
    & \times \sum_{m=-j}^{j} \; \sum_{j'=(n-2k-z) \downarrow 2} \; \sum_{m'=-j'}^{j'} \; \sum_{j''=z \downarrow 2} \; \sum_{m''=-j''}^{j''} \; \sum_{\lambda=0}^{\ell} \; \sum_{\lambda'=0}^{\ell-\lambda} \nonumber \\
    & \times \sum_{\mu = - \lambda}^{\lambda} \; \sum_{\mu' = -\lambda'}^{\lambda'} \; \sum_{L_0 M_0} \; \sum_{\mathcal{L}_0 \mathcal{M}_0} \; \sum_{L_1 M_1} \; \sum_{\mathcal{L}_1 \mathcal{M}_1} \; \sum_{L_2 M_2} \; \sum_{\mathcal{L}_2 \mathcal{M}_2} \nonumber \\ \nonumber \\ \nonumber
    & \times \mathcal{U}_{z z' z'' j j' j''}^{k n \;(\ell/2)}  \; Q_{\ell \lambda \lambda'} \; \mathcal{G}_{\lambda \lambda' j j' j'' L_0 \mathcal{L}_0 L_1 \mathcal{L}_1 \ell L_2 \mathcal{L}_2}^{\mu \mu' m m'm'' M_0 \mathcal{M}_0 M_1 \mathcal{M}_1 M_2 \mathcal{M}_2} \nonumber \\ \nonumber \\
    & \times \epsilon_{12,1}^{n-z-z'-2z''+\lambda'} \; \epsilon_{13,1}^{z+z'+2z''+\ell-\lambda-\lambda'} \\ \nonumber \\ \nonumber
    & \times Y_{L_0}^{M_0}(\hat{x}_1) \; Y_{L_1}^{M_1}(\hat{r}_{12}) \;  Y_{L_2}^{M_2}(\hat{r}_{13}) , \\ \nonumber
\end{align}
where we have defined
\begin{align}
    \mathcal{G}&_{\lambda \lambda' j j' j'' L_0 \mathcal{L}_0 L_1 \mathcal{L}_1 \ell L_2 \mathcal{L}_2}^{\mu \mu' m m'm'' M_0 \mathcal{M}_0 M_1 \mathcal{M}_1 M_2 \mathcal{M}_2} \equiv \\
    & \hspace{2cm} \mathcal{G}_{\lambda j' \mathcal{L}_0}^{\mu -m' -\mathcal{M}_0} \; \mathcal{G}_{\mathcal{L}_0 j'' L_0}^{\mathcal{M}_0 -m'' -M_0} \; \mathcal{G}_{\lambda' j \mathcal{L}_1}^{\mu' m -\mathcal{M}_1} \nonumber \\
    & \hspace{2cm} \times \mathcal{G}_{\mathcal{L}_1 j' L_1}^{\mathcal{M}_1 m' -M_1} \; \mathcal{G}_{(\ell-\lambda-\lambda') j \mathcal{L}_2}^{(m-\mu-\mu') -m -\mathcal{M}_2} \; \mathcal{G}_{\mathcal{L}_2 j'' L_2}^{\mathcal{M}_2 m'' -M_2}. \nonumber 
    \end{align}
$\mathcal{G}_{\ell_1 \ell_2 \ell_3}^{m_1 m_2 m_3}$ is a Gaunt integral, defined in equation (\ref{eqn:gaunt}).

\begin{table*}
\caption{List of variables being summed over in our calculation, with their ranges and explanations.}
\centering
\begin{tabular}{ |p{1cm}||p{1cm}|p{1cm}|p{11cm}|  }
 \hline
 \multicolumn{4}{|c|}{Detailed Summation Limits} \\
 \hline
 Variable& Min. & Max. & Explanation\\
 \hline
 $n$ & $0$ & $\infty$ & From the expansion of $a^{-\ell} (1-2xt+t^2)^{-\alpha}$ into Gegenbauer polynomials.\\
 $k$ & $0$ & $\lfloor n/2 \rfloor$ &  From the explicit form for the Gegenbauer polynomial $C_n^\alpha$.\\
 $z$ & $0$ & $n-2k$ & From the expansion of $(\epsilon_{12,1} \hat{r}_{12} \cdot \hat{x}_1 + \epsilon_{13,1} \hat{r}_{13} \cdot \hat{x}_1)$ using the binomial theorem.\\
 $z'$ & $0$ & $k$ & From the first time we expanded $(\epsilon_{12,1}^2 + \epsilon_{13,1}^2 + 2r_{12}r_{13}(\hat{r}_{12}\cdot\hat{r}_{13}))^k$ using the binomial theorem, treating the sum of squares as the first term, and the rest as the second one.\\
 $z''$ & $0$ & $k-z'$ & From the second time we expanded $[\epsilon_{12,1}^2 + \epsilon_{13,1}^2 + 2r_{12}r_{13}(\hat{r}_{12}\cdot\hat{r}_{13})]^k$ using the binomial theorem, which after the first expansion has the term $(\epsilon_{12,1}^2 + \epsilon_{13,1}^2)k-z'$, which we now further expand.\\
 $j$ & z' & 0 & Sum ranging from $z'$ to zero in steps of two, because $\hat{r}_{12} \cdot \hat{r}_{13}$ is to the power of $z'$. This comes from expanding the power in terms of Legendre polynomials.\\
 $m$ & $-j$ & $j$ & Sum over spherical harmonics after using the addition theorem for the Legendre polynomial of $\hat{r}_{12} \cdot \hat{r}_{13}$ \\
 $j'$ & $n-2k-z$ & $0$ & Sum ranging from $n-2k-z$ to zero in steps of two, because $\hat{r}_{12} \cdot \hat{x}_{1}$ is to the power of $n-2k-z$. This comes from expanding the power in terms of Legendre polynomials. \\
 $m'$ & $-j'$ & $j'$ & Sum over spherical harmonics after using the addition theorem for the Legendre polynomial of $\hat{r}_{12} \cdot \hat{x}_{1}$. \\
 $j''$ & $z$ & $0$ & Sum ranging from $z$ to zero in steps of two, because $\hat{r}_{13} \cdot \hat{x}_{1}$ is to the power of $z$. This comes from expanding the power in terms of Legendre polynomials. \\
 $m''$ & $-j''$ & $j''$ & Sum over spherical harmonics after using the addition theorem for the Legendre polynomial of $\hat{r}_{13} \cdot \hat{x}_{1}$. \\
 $\lambda$ & 0 & $\ell$ & First sum of the solid harmonic shift theorem applied to $3\vec{x}_1+\vec{d}$. \\
 $\mu$ & $-\lambda$ & $\lambda$ & Second sum of the solid harmonic shift theorem applied to $3\vec{x}_1+\vec{d}$. \\
 $\lambda'$ & $0$ & $\ell-\lambda$ & First sum of the solid harmonic shift theorem applied when we open up $d$ as $\vec{r}_{12}+\vec{r}_{13}$. \\
 $\mu'$ & $-\lambda'$ & $\lambda'$ & Second sum of the solid harmonic shift theorem applied when we open up $d$ as $\vec{r}_{12}+\vec{r}_{13}$. \\
 \hline
\label{table1}
\end{tabular}
\end{table*}

\section{Cartesian Proof of Low-$\ell$ Cases of Solid Harmonic Shift Theorem}\label{app:app2}
Here we will use explicit calculation to prove the solid harmonic shift theorem used in Appendix \ref{app:app1} for $\ell = \{0,1,2\}$. For a vector $\vec{r}$, its solid harmonic form is:
\begin{align}
    R_\ell^m (\vec{r}) \equiv \sqrt{\frac{4\pi}{2\ell+1}} r^\ell Y_{\ell}^{m} (\hat{r})
\end{align}
Table \ref{table2} shows $Y_\ell^m(\hat{r})$ and $R_\ell^m(\vec{r})$ for different values of $\ell$ and $m$. We are interested in finding the solid harmonics of a sum of vectors as a function of their solid harmonics individually. We will then find out what $R_\ell^m (\vec{r}+\vec{s})$ is.
\\

\noindent \underline{For $\ell=0$ and $m=0$:}
\begin{align}
    R_0^0(\vec{r} + \vec{s}) = \frac{1}{2} \left[ R_0^0(\vec{r}) + R_0^0(\vec{s}) \right] = 1 .
\end{align}
\vspace{0.1mm}

\noindent \underline{For $\ell=1$ and $m=0$:}
\begin{align}
    R_1^0(\vec{r} + \vec{s}) = r_z + s_z .
\end{align}
\vspace{0.1mm}

\noindent \underline{For $\ell=1$ and $m=1$:}
\begin{align}
    \label{eq:R10}
    R_1^1(\vec{r} + \vec{s}) &= -\frac{\sqrt{2}}{2} [r_x + s_x + i(r_y+s_y)] \\
    &= -\frac{\sqrt{2}}{2} (r_x + ir_y) - \frac{\sqrt{2}}{2} (s_x + is_y)
    \nonumber
\end{align}
Here, $-(\sqrt{2}/2) (r_x + ir_y)$ and $-(\sqrt{2}/2) (s_x + is_y)$ are $R_1^1(\vec{r})$ and $R_1^1(\vec{s})$ (see Table \ref{table2}), respectively. So we find
\begin{align}
    \label{eq:R11}
    R_1^1(\vec{r} + \vec{s}) &= R_1^1(\vec{r}) + R_1^1(\vec{s}) .
\end{align}
\vspace{0.1mm}

\noindent \underline{For $\ell=1$ and $m=-1$:}

\noindent Following the same logic, for $\ell=1$ and $m=-1$ we find that
\begin{align}
    \label{eq:R11_2}
    R_1^{-1}(\vec{r} + \vec{s}) &= R_1^{-1}(\vec{r}) + R_1^{-1}(\vec{s}) .
\end{align}
\vspace{0.1mm}

\noindent \underline{For $\ell=2$ and $m=0$:}
\begin{align}
    R_2^0 (\vec{r}+\vec{s}) = -\frac{1}{2}[2(r_z + s_z)^2 -(r_x+s_x)^2 - (r_y+s_y)^2]
\end{align}
which we find after expanding each term
\begin{align}
    \label{eqn:R20}
    R_2^0 (\vec{r}+\vec{s}) &= R_2^0(\vec{r}) + R_2^0(\vec{s}) + 2 r_z s_z - r_x s_x - r_y s_y \\
    &= R_2^0(\vec{r}) + R_2^0(\vec{s}) + 2 R_1^0(\vec{r}) R_1^0(\vec{s}) - r_x s_x - r_y s_y
    \nonumber
\end{align}
We know from the derivation of $R_1^1(\vec{v})$ and $R_1^{-1}(\vec{v})$ above that
\begin{align}
    v_x &= \frac{1}{\sqrt{2}} [R_1^{-1}(\vec{v})-R_1^1(\vec{v})] \\
    v_y &= -\frac{1}{i \sqrt{2}} [R_1^{1}(\vec{v})+R_1^{-1}(\vec{v})]
\end{align}
After substituting these into Eq. (\ref{eqn:R20}), we find that
\begin{align}
    R_2^0 (\vec{r}+\vec{s}) &= R_2^0(\vec{r}) + R_2^0(\vec{s}) + 2 R_1^0(\vec{r}) R_1^0(\vec{s}) \nonumber \\
    &- \frac{1}{2} \big[ R_1^{-1}(\vec{r}) R_1^{-1}(\vec{s}) + R_1^{-1}(\vec{r}) R_1^{1}(\vec{s}) \nonumber \\
    &+ R_1^{1}(\vec{r}) R_1^{-1}(\vec{s}) + R_1^{1}(\vec{r}) R_1^{1}(\vec{s}) \big] \\
    &- \frac{1}{2} \big[ R_1^{1}(\vec{r}) R_1^{1}(\vec{s}) + R_1^{1}(\vec{r}) R_1^{-1}(\vec{s}) \nonumber \\
    &+ R_1^{-1}(\vec{r}) R_1^{1}(\vec{s}) + R_1^{-1}(\vec{r}) R_1^{-1}(\vec{s}) \big] ,
    \nonumber
\end{align}
which simplifying becomes
\begin{align}
    R_2^0 (\vec{r}+\vec{s}) &= R_2^0(\vec{r}) + R_2^0(\vec{s}) + 2 R_1^0(\vec{r}) R_1^0(\vec{s}) \\
    &+ R_1^{1}(\vec{r}) R_1^{-1}(\vec{s}) + R_1^{-1}(\vec{r}) R_1^{1}(\vec{s})
    \nonumber
\end{align}

\vspace{0.1mm}

\noindent \underline{For $\ell=2$ and $m=1$:}
\begin{align}
    R_2^1(\vec{r}&+\vec{s}) = -\frac{\sqrt{6}}{2} \left[ r_x+s_x+i(r_y+s_y) \right](r_z+s_z) \nonumber \\
    &= - \frac{\sqrt{6}}{2}(r_x+ir_y)r_z - \frac{\sqrt{6}}{2}(s_x+is_y)s_z \\
    &\hspace{3mm}- \frac{\sqrt{6}}{2}(r_x+ir_y)s_z - \frac{\sqrt{6}}{2}(s_x+is_y)r_z \nonumber \\
    &= R_2^1(\vec{r}) + R_2^1(\vec{s}) + \sqrt{3} \left[ R_1^1(\vec{r})R_1^0(\vec{s}) + R_1^1(\vec{s})R_1^0(\vec{r}) \right]
    \nonumber
\end{align}
\vspace{0.1mm}

\noindent \underline{For $\ell=2$ and $m=-1$:}
\begin{align}
    R_2^1(\vec{r}&+\vec{s}) = \frac{\sqrt{6}}{2} \left[ r_x+s_x-i(r_y+s_y) \right](r_z+s_z) \nonumber \\
    &= \frac{\sqrt{6}}{2}(r_x-ir_y)r_z + \frac{\sqrt{6}}{2}(s_x-is_y)s_z \\
    &\hspace{3mm} \frac{\sqrt{6}}{2}(r_x-ir_y)s_z + \frac{\sqrt{6}}{2}(s_x-is_y)r_z \nonumber \\
    &= R_2^{-1}(\vec{r}) + R_2^{-1}(\vec{s}) + \sqrt{3} \left[ R_1^{-1}(\vec{r})R_1^0(\vec{s}) + R_1^{-1}(\vec{s})R_1^0(\vec{r}) \right]
    \nonumber
\end{align}
\vspace{0.1mm}

\noindent \underline{For $\ell=2$ and $m=2$:}
\begin{align}
    R_2^2(\vec{r}&+\vec{s}) = \frac{\sqrt{6}}{4} (r_x+ir_y + s_x+is_y)^2  \\
    &= \frac{\sqrt{6}}{4} \left[ (r_x+ir_y)^2 + 2(r_x+ir_y)(s_x+is_y) + (s_x+is_y)^2\right] \nonumber\\
    &= R_2^2(\vec{r})+R_2^2(\vec{s})+\sqrt{6} R_1^1(\vec{r}) R_1^1(\vec{s})\nonumber
\end{align}
\vspace{0.1mm}

\noindent \underline{For $\ell=2$ and $m=-2$:}
\begin{align}
    R_2^2(\vec{r}&+\vec{s}) = \frac{\sqrt{6}}{4} (r_x-ir_y + s_x-is_y)^2 \\
    &= \frac{\sqrt{6}}{4} \left[ (r_x-ir_y)^2 + 2(r_x-ir_y)(s_x-is_y) + (s_x-is_y)^2\right] \nonumber\\
    &= R_2^{-2}(\vec{r})+R_2^{-2}(\vec{s})+\sqrt{6} R_1^{-1}(\vec{r}) R_1^{-1}(\vec{s})\nonumber
\end{align}

\begin{table*}
\caption{Spherical and solid harmonics of $\hat{r}$ and $\vec{r}$, respectively, for $\ell=\{0, 1, 2\}$.}
\centering
\large
\begin{tabular}{ |p{2cm}| |p{2cm}|p{3.5cm}|p{4cm}|  }
 \hline
 \multicolumn{4}{|c|}{Solid Harmonics} \\
 \hline
 $\ell$ & $m$ & $Y_\ell^m(\hat{r})$ & $R_\ell^m(\vec{r}) = \sqrt{\frac{4\pi}{2\ell+1}}r^\ell Y_\ell^m (\hat{r})$\\
 \hline
 \vspace{2mm} 0 & \vspace{2mm} 0 & \vspace{2mm} $\frac{1}{2} \sqrt{\frac{1}{\pi}}$ & \vspace{2mm} 1 \\
 \vspace{2mm} 1 & \vspace{2mm} -1 & \vspace{2mm} $\frac{1}{2} \sqrt{\frac{3}{2\pi}} (r_x-ir_y) \; r^{-1}$ & \vspace{2mm} $\frac{\sqrt{2}}{2}(r_x-ir_y)$ \\
 \vspace{2mm} 1 & \vspace{2mm} 0 & \vspace{2mm} $\frac{1}{2} \sqrt{\frac{3}{\pi}} r_z \; r^{-1}$ & \vspace{2mm} $r_z$ \\
 \vspace{2mm} 1 & \vspace{2mm} 1 & \vspace{2mm} $-\frac{1}{2} \sqrt{\frac{3}{2\pi}} (r_x+ir_y) \; r^{-1}$ & \vspace{2mm} $-\frac{\sqrt{2}}{2}(r_x+ir_y)$ \\
 \vspace{2mm} 2 & \vspace{2mm} -2 & \vspace{2mm} $\frac{1}{4} \sqrt{\frac{15}{2\pi}} (r_x-ir_y)^2 \; r^{-2}$ & \vspace{2mm} $\frac{\sqrt{6}}{4} (r_x-ir_y)^2$ \\
 \vspace{2mm} 2 & \vspace{2mm} -1 & \vspace{2mm} $\frac{1}{2} \sqrt{\frac{15}{2\pi}} (r_x-ir_y) \; z \; r^{-2}$ & \vspace{2mm} $\frac{\sqrt{6}}{2} (r_x-ir_y)\;r_z$ \\
 \vspace{2mm} 2 & \vspace{2mm} 0 & \vspace{2mm} $\frac{1}{4} \sqrt{\frac{5}{\pi}} (2r_z^2-r_x^2-r_y^2) \; r^{-2}$ & \vspace{2mm} $\frac{1}{2} (2r_z^2-r_x^2-r_y^2)$ \\
 \vspace{2mm} 2 & \vspace{2mm} 1 & \vspace{2mm} $-\frac{1}{2} \sqrt{\frac{15}{2\pi}} (r_x+ir_y) \; z \; r^{-2}$ & \vspace{2mm} $- \frac{\sqrt{6}}{2} (r_x+ir_y)\;r_z$ \\
 \vspace{2mm} 2 & \vspace{2mm} 2 & \vspace{2mm} $\frac{1}{4} \sqrt{\frac{15}{2\pi}} (r_x+ir_y)^2 \; r^{-2}$ & \vspace{2mm} $\frac{\sqrt{6}}{4} (r_x+ir_y)^2$ \\
 \hline
\label{table2}
\end{tabular}
\end{table*}

\section{Product to Sum Expansions for Spherical Harmonics}
\label{app:app3}
Here we show how a product of three spherical harmonics of the same argument can be turned into a sum over single harmonics. We start with the product of two spherical harmonics ($Y_{l_1}^{m_1}(\hat{x})$ and $Y_{l_2}^{m_2}(\hat{x})$), which can be written as a sum of other spherical harmonics times a coefficient $C_{l_1 l_2 L}^{m_1 m_2 M}$ as in
\begin{align}
    \label{eqn:app3_1}
    Y_{l_1}^{m_1}(\hat{x}) \;  Y_{l_2}^{m_2}(\hat{x}) = \sum_{LM} C_{l_1 l_2 L}^{m_1 m_2 M} \; Y_L^M (\hat{x}) .
\end{align}
Multiplying both sides by $\int d\Omega_x Y_L^{M*}(\hat{x})$ we obtain:
\begin{align}
    \int d\Omega_x Y_{l_1}^{m_1}(\hat{x}) \; Y_{l_2}^{m_2}(\hat{x}) \; Y_L^{M*}(\hat{x}) = C_{l_1 l_2 L}^{m_1 m_2 M} .
\end{align}
We know that
\begin{align}
    Y_L^{M*} (\hat{x}) = (-1)^M Y_L^{-M} (\hat{x}) .
\end{align}
Thus we find
\begin{align}
    \label{eqn:C}
    (-1)^M \int d\Omega_x Y_{l_1}^{m_1}(\hat{x}) \; Y_{l_2}^{m_2}(\hat{x}) \; Y_L^{-M}(\hat{x}) = C_{l_1 l_2 L}^{m_1 m_2 M}
\end{align}
The integral above is the Gaunt integral, which we denote
\begin{align}
    \label{eqn:gaunt}
    \mathcal{G}_{l_1 l_2 L}^{m_1 m_2 -M} = &\sqrt{\frac{(2l_1+1)(2l_2+2)(2l_3+3)}{4\pi}} \nonumber \\
    & \times \begin{pmatrix}
    l_1 & l_2 & L\\ 
    0 & 0 & 0
    \end{pmatrix}
    \begin{pmatrix}
    l_1 & l_2 & L\\ 
    m_1 & m_2 & -M
    \end{pmatrix} .
\end{align}
Substituting equation (\ref{eqn:gaunt}) into equation (\ref{eqn:C}) we obtain
\begin{align}
    \label{eqn:app3_2}
    C_{l_1 l_2 L}^{m_1 m_2 M} = (-1)^M \mathcal{G}_{l_1 l_2 L}^{m_1 m_2 -M} .
\end{align}

Now we want to extend this result to the product of three spherical harmonics, i.e.
\begin{align}
    \label{eqn:three}
    Y_{l_1}^{m_1}(\hat{x}) Y_{l_2}^{m_2}(\hat{x}) Y_{l_3}^{m_3}(\hat{x}) = \sum_{LM} D_{l_1 l_2 l_3 L}^{m_1 m_2 m_3 M} \; Y_L^M (\hat{x}) .
\end{align}
Multiplying both sides of equation (\ref{eqn:three}) by $\int d\Omega_x Y_L^{M*}(\hat{x})$ we find:
\begin{align}
    \int d\Omega_x Y_{l_1}^{m_1}(\hat{x}) \; Y_{l_2}^{m_2}(\hat{x}) \; Y_{l_3}^{m_3}(\hat{x}) \; Y_L^{M*}(\hat{x}) = D_{l_1 l_2 l_3 L}^{m_1 m_2 m_3 M}
\end{align}
Transforming the conjugate using $Y_L^{M*} (x) = (-1)^M Y_L^{-M} (x)$ we obtain:
\begin{align}
    \label{eqn:app3_4}
    (-1)^{M} \int d\Omega_x Y_{l_1}^{m_1}(\hat{x}) \; Y_{l_2}^{m_2}(\hat{x}) \; Y_{l_3}^{m_3}(\hat{x}) \; Y_L^{-M}(\hat{x}) = D_{l_1 l_2 l_3 L}^{m_1 m_2 m_3 M}
\end{align}
From equation (\ref{eqn:app3_1}) we see that
\begin{align}
    \label{eqn:app3_5}
    Y_{l_1}^{m_1}(\hat{x}) \; Y_{l_2}^{m_2}(\hat{x}) = \sum_{\mathcal{L} \mathcal{M}} C_{l_1 l_2 \mathcal{L}}^{m_1 m_2 \mathcal{M}} \; Y_\mathcal{L}^\mathcal{M} (\hat{x})
\end{align}
Thus, if we substitute equation (\ref{eqn:app3_5}) in equation (\ref{eqn:app3_4}) we obtain
\begin{align}
    \int d\Omega_x  \; \sum_{\mathcal{L} \mathcal{M}} &(-1)^{M+\mathcal{M}} \; C_{l_1 l_2 \mathcal{L}}^{m_1 m_2 \mathcal{M}} \\
    &\times Y_\mathcal{L}^\mathcal{M} (\hat{x})\; Y_{l_3}^{m_3}(\hat{x}) \; Y_L^{-M}(\hat{x}) = D_{l_1 l_2 l_3 L}^{m_1 m_2 m_3 M} , \nonumber
\end{align}
which based on equation (\ref{eqn:gaunt}) we know is
\begin{align}
    \label{eqn:app3_7}
    \sum_{\mathcal{L} \mathcal{M}} (-1)^{M+\mathcal{M}} \; C_{l_1 l_2 \mathcal{L}}^{m_1 m_2 \mathcal{M}} \; \mathcal{G}_{\mathcal{L} l_3 L}^{\mathcal{M} m_3 -M} = D_{l_1 l_2 l_3 L}^{m_1 m_2 m_3 M} .
\end{align}
And by substituting equation (\ref{eqn:app3_2}) in equation (\ref{eqn:app3_7}) we find
\begin{align}
    \label{eqn:app3_almostthere}
    D_{l_1 l_2 l_3 L}^{m_1 m_2 m_3 M} = \sum_{\mathcal{L} \mathcal{M}} (-1)^{M+\mathcal{M}} \; \mathcal{G}_{l_1 l_2 \mathcal{L}}^{m_1 m_2 -\mathcal{M}} \; \mathcal{G}_{\mathcal{L} l_3 L}^{\mathcal{M} m_3 -M} .
\end{align}
Finally, substituting equation (\ref{eqn:app3_almostthere}) in equation (\ref{eqn:three}) we obtain
\begin{align}
    Y_{l_1}^{m_1}(\hat{x}) Y_{l_2}^{m_2}&(\hat{x}) Y_{l_3}^{m_3}(\hat{x}) =\\ &\sum_{LM} \sum_{\mathcal{L}\mathcal{M}} (-1)^{M+\mathcal{M}} \mathcal{G}_{l_1 l_2 \mathcal{L}}^{m_1 m_2 -\mathcal{M}} \; \mathcal{G}_{\mathcal{L} l_3 L}^{\mathcal{M} m_3 -M} \; Y_L^M (\hat{x}) . \nonumber
\end{align}

\bibliographystyle{mnras}
\bibliography{3pcf_los_bib}

%\begin{thebibliography}{99}

%\end{thebibliography}

%%%%%%%%%%%%%%%%%%%%%%%%%%%%%%%%%%%%%%%%%%%%%%%%%%

% Don't change these lines
\bsp	% typesetting comment
\label{lastpage}
\end{document}